\documentclass[aps,showpacs,eqsecnum,preprint,amsmath,amssymb,floats,epsf,nofootinbib]{revtex4}

\usepackage{epsfig}
\usepackage{graphicx}
\usepackage{dcolumn}
\usepackage{bm}

\def\DESepsf(#1 width #2){\epsfxsize=#2 \epsfbox{#1}}

\def\lsim{ {\ \lower-1.2pt\vbox{\hbox{\rlap{$<$}\lower5pt\vbox{\hbox{$\sim$}
}}}\ } }
\def\gsim{ {\ \lower-1.2pt\vbox{\hbox{\rlap{$>$}\lower5pt\vbox{\hbox{$\sim$}
}}}\ } }

\begin{document}

\preprint{hep-ph/0509092}

\title{\vspace*{1cm} Weak Annihilation Topologies and Final State Interactions in $D \to PP$ Decays}

\author{\vspace*{0.5cm} \normalsize \sc Jr-Hau Lai}
\author{\normalsize \sc Kwei-Chou Yang}

\affiliation{\vspace*{0.3cm} \normalsize\sl Department of Physics,
Chung Yuan Christian University, Chung-Li, Taiwan 320, Republic of
China \vspace*{2cm}}


\small
\begin{abstract}
We study two-body $D\to PP$ decays, assuming that each decay
process go through the {\it bare} amplitude followed by elastic
SU(3) rescattering, where the {\it bare} amplitude consists of (i)
the color-allowed and color-suppressed factorization amplitudes
and (ii) the short-distance weak annihilation amplitudes. We have
performed the $\chi^2$ fit on 14 branching ratios of $D\rightarrow
PP$ decays in the formalism of the above mentioned model. The
final state interactions can be well accounted for by the
short-distance annihilation topologies and SU(3) rescatterings.
The two SU(3) rescattering phase differences are $\delta\equiv
\delta_{27}-\delta_{8}\simeq -46^\circ$ and $\sigma\equiv
\delta_{27}-\delta_{1}\simeq -21^\circ$, where $\delta_{27},
\delta_{8}$ and $\delta_{1}$ are the rescattering phases of final
states corresponding to the representations $\mathbf{27},
\mathbf{8}$ and $\mathbf{1}$, respectively. We find that the $D^0
\to K^0 \overline K^0$ decay occurs mainly due to the nonzero
short-distance weak annihilation effects, originating from SU(3)
symmetry-breaking corrections to the distribution amplitudes of
the final-state kaons, but receives tiny effects from other modes
via SU(3) rescattering. Our results are in remarkable accordance
with the current data.

\end{abstract}



\maketitle

\section{Introduction}\label{sec:introduction}

 It is known that the na\"{\i}ve factorization approximation fails
to describe the color-suppressed $D$ decays. The results can be
improved if the Fierz-transformed terms characterized by $1/N_c$
are discarded~\cite{Buras:1985xv}. The short-distance (SD) weak
annihilation effects, which may mimic some non-resonant final
state interactions, have recently been emphasized in two-body $B$
decays~\cite{Yeh:1997rq,KLS00,Beneke:2001ev,Beneke:2003zv}. In $D$
decays, the SD weak annihilation contributions involving gluon
emission from the final-state quarks, which arise from the
$(V-A)\otimes (V-A)$ four-quark operators, vanish. Nevertheless,
if the gluon is emitted from the initial quarks, the SD weak
annihilation effects are not zero (see the results shown in
Sec.~\ref{subsec:ann}) and may give sizable corrections to the
amplitudes. Such effects were first noticed by Li and
Yeh~\cite{Yeh:1997rq,KLS00} and recently discussed in $B$
decays~\cite{Beneke:2001ev,Beneke:2003zv}. One therefore expects
that the SD weak annihilation may play an important role in $D$
decays because the energy released to the final-state particles is
not as large as that in $B$ decays. Unfortunately, the SD weak
annihilation topologies, in general, are not calculable in the QCD
factorization approach\footnote{One may introduce the transverse
momenta of quarks (${\bf k}_\perp$) to regulate the end-point
divergence, where ${\bf k}_\perp$ is naturally constrained by the
infrared cutoff $\sim 1/R$ with $R$ the meson's radius (some other
discussions can be found in Ref.~\cite{Beneke:2001ev}). However,
the result may suffer from the gauge problem and is part of higher
twist contribution. It is interesting to note that in deeply
inelastic scattering (DIS) processes, by introducing a generalized
special propagator~\cite{Qiu:1988dn} for massive
quarks~\cite{Yang:1997er}, the separation of the hard part ($T$)
from the soft part (parton distributions) is manifestly gauge
invariant for different orders in $1/Q$ (twist). An important
feature of using the special propagator technique is that the
${\bf k}_\perp$ contributions should be moved into $T$, such that,
after combining with the gluon field $A^\alpha$ in $T$, a
covariant derivative of color gauge invariance can be achieved and
classified as a high-twist contribution.}.

The color-suppressed $\overline{B}^{0}\to D^{(*) 0} (\pi^{0},
\eta, \omega) $,  $\overline{B}^{0}\to D^{0} (\eta',\overline
K^0)$  and $B^- \to D_s K^-$ decay modes have recently been
observed by the Belle, CLEO and Barbar
collaborations~\cite{Abe:2001zi,Coan:2001ei,Krokovny:2002pe,Aubert:2002vg,Schumann:2005ej}.
These branching ratios (BRs) are much larger than the expectation
in the factorization-based analysis~\cite{Cheng:1998kd}. Using the
isospin amplitude analysis of $\overline{B}^{0}\rightarrow
D^{0}\pi^{-}$, $\overline{B}^{0}\rightarrow D^{+}\pi^{-}$ and
$\overline{B}^{0}\rightarrow D^{0}\pi^{0} $, one can obtain that
the rescattering phase difference of isospin amplitudes $A_{3/2}$
and $A_{1/2}$ is about $30^{\circ }$~\cite{Yang}. It may indicate
long-distance (LD) final state interactions (FSIs) are not
negligible even in $B$ meson decays~\cite{Yang}. Analogously,
larger FSIs could be expected in $D$ meson decays since the energy
released in $D$ decays is much less than that in $B$ decays as
mentioned above. For illustrating this point, we perform the
isospin decomposition for $D^{0}\rightarrow K^{-}\pi^{+},
\overline{K}^{0}\pi^{0}$ and $D^{+}\rightarrow
\overline{K}^{0}\pi^{+}$ decay amplitudes:
\begin{eqnarray}
A (D^{0}\rightarrow K^{-}\pi^{+} )
 &=&\sqrt{\frac{1}{3}}A_{3/2}+ \sqrt{\frac{2}{3}}A_{1/2},  \nonumber \\
A ( D^{0}\rightarrow \overline{K}^{0}\pi^{0} )
 &=&
 \sqrt{\frac{2}{3}}A_{3/2}-\sqrt{\frac{1}{3}}A_{1/2},  \nonumber \\
A ( D^{+}\rightarrow \overline{K}^{0}\pi^{+} )
&=&\sqrt{3}A_{3/2},\label{eq:su2}
\end{eqnarray}
where the isospin amplitudes with isospin $3/2$ and $1/2$ are
denoted as $A_{3/2}$ and $A_{1/2}$, respectively. The relative
rescattering phase between $A_{3/2}$ and $A_{1/2}$, denoted as
$\phi $, satisfies the following relation,
\begin{equation}
\cos \phi =\frac{ |A ( D^{0}\rightarrow K^{-}\pi^{+} ) |^{2}-2 | A
( D^{0}\rightarrow \overline{K}^{0}\pi^{0} ) |^{2}+\frac{1}{3} | A
( D^{+}\rightarrow \overline{K}^{0}\pi^{+} ) |^{2}}{2\sqrt{2}
|A_{1/2}| \ | A_{3/2} | }. \label{eq:su2phase}
\end{equation}
Substituting the data for BRs of $D^{0}\rightarrow K^{-}\pi^{+},
\overline{K}^{0}\pi^{0}$ and $D^{+}\rightarrow \overline{K}^{0}\pi
^{+}$ modes, which are $(3.8\pm 0.09) \%$, $(2.30 \pm 0.22) \%$
and $(2.82\pm 0.19) \%$~\cite{PDG}, respectively, into
Eq.~(\ref{eq:su2phase}), one can obtain the rescattering phase
$\phi \approx 94^{\circ }$, much larger than that in the charmful
two-body $B$ decays. The above result indicates that FSIs should
be significant in $D$ meson decays.

In this article, we will assume that the FSIs in $D\to PP$ are
described by SD weak annihilation topologies and elastic (LD)
SU(3) rescatterings. Analogously, the elastic final-state
rescattering picture has been extended from SU(2)-type to
SU(3)-type in $B$ decays~\cite{Yang,Chua:2002wk}. We presume that
each $D\to PP$ decay process go through the ``{\it bare}"
amplitude followed by elastic-SU(3) rescattering, where the {\it
bare} amplitude describing the SD-dominant contributions consists
of (i) the usual factorization amplitudes of color-allowance and
color-suppression, which can be calculated using the factorization
approach, and (ii) the SD weak annihilation topologies
($W$-exchange or $W$-annihilation) which present the endpoint
singularities are regulated by introducing the complex
phenomenological parameter $X_A$~\cite{Beneke:2001ev} in the QCD
factorization approach (see the detailed description in
Sec.~\ref{subsec:ann}).

Interestingly, the SD weak annihilation amplitudes are dominated
by the topologies of gluon emission arising from the {\it
initial-state quarks} of the weak vertex, while the total
amplitudes vanish in order of $\alpha_s$ if the gluon is emitted
from the {\it final-state quarks}. On the hand hand, the elastic
SU(3) rescatterings are mainly generated by gluon exchange between
the final-state mesons. Therefore, it could be expected that the
possible double counting is negligible between the two possible
sources for FSIs. We will give a detailed discussion for possible
rescattering sources in Sec.~\ref{sec:summary}.

 We consider the SU(3) breaking effects in the {\it bare}
amplitude level, but, for simplicity, do not distinguish the
breaking influence on the two SU(3) rescattering phases, defined
as $\delta\equiv \delta_{27}-\delta_{8}$ and $\sigma\equiv
\delta_{27}-\delta_{1}$. In other words, in description of decay
amplitudes, masses vary according to SU(3) breaking, and meson
productions differ in strength as reflected in the decay constants
and form factors.

The rest of this article is organized as follows. In
Sec.~\ref{sec:bar_amp}, neglecting elastic SU(3) FSIs, we first
sketch the factorization amplitudes as well as the SD weak
annihilation contributions in two-body $D$ decays.
Sec.~\ref{sec:fsi} is devoted to the formulation of SU(3)
rescatterings. We give the numerical analysis in
Sec.~\ref{sec:results}. The discussions and summary are presented
in Sec.~\ref{sec:summary}. The detailed results for the
factorization amplitudes, SD weak annihilation amplitudes and
tensor approach for the SU(3) final-state decomposition are
collected in Appendices~\ref{app:fac}, \ref{app:ann}, and
\ref{appsec:fsi}, respectively.

\section{The {\it bare} amplitudes}\label{sec:bar_amp}

Here we present factorization and SD weak annihilation amplitudes
for $D \to P P$ decays. The relevant effective Hamiltonian for the
charmed meson decays is
\begin{equation}
{\cal H}_{\rm eff}=\frac{G_{F}}{\sqrt{2}}
\sum_{q,q^{\prime}=d,s}V_{uq}V_{cq^{\prime }}^{\ast }\left(
c_{1}O_{1}+ c_{2}O_{2}\right) +H.c.,
\end{equation}
where $G_{F}$ is the weak coupling constant, and the
current-current operators read
\begin{equation}
O_{1}=\left( \overline{u}q\right)_{V-A}\left(
\overline{q}^{\prime}c\right)_{V-A},\ \ O_{2}=\left( \overline{u}%
c\right)_{V-A}\left( \overline{q}^{\prime }q\right)_{V-A},
\end{equation}
with $\left( \overline{u}q\right)_{V-A}\equiv
\overline{u}\gamma^{\mu }(1-\gamma^{5})q$. $V_{uq}$ and
$V_{cq^{\prime }}^{\ast}$ are the Cabibbo-Kobayashi-Maskawa (CKM)
matrix elements given by
\begin{equation}
\left(
\begin{array}{ccc}
V_{ud} & V_{us} & V_{ub} \\
V_{cd} & V_{cs} & V_{cb} \\
V_{td} & V_{ts} & V_{tb}
\end{array}
\right) =\left(
\begin{array}{ccc}
1-\frac{\lambda^{2}}{2} & \lambda & A\lambda^{2}(\rho -i\eta ) \\
-\lambda & 1-\frac{\lambda^{2}}{2} & A\lambda^{2} \\
A\lambda^{3}(1-\rho -i\eta ) & -A\lambda^{2} & 1
\end{array}
\right)
\end{equation}
in the Wolfenstein parametrization.

The two ingredients of the ``{\it bare}'' amplitude
for describing the decay processes are (i) the factorization
amplitudes, which are made of the color-allowed external
$W$-emission tree amplitude (${\cal T}$) and/or the
color-suppressed internal $W$-emission amplitude (${\cal C}$), and
(ii) the weak annihilation amplitudes which consist of
$W$-exchange and/or $W$-annihilation topologies.

\subsection{Factorization amplitudes}\label{subsec:FacAmp}

Taking $D^0 \to K^- \pi^+, \overline K^0 \pi^0$ as examples, the
factorization amplitudes can be written as the following general
forms:
\begin{eqnarray}
{\cal T}^{K^- \pi^+}
&=&\frac{G_{F}}{\sqrt{2}}V_{ud}V_{cs}^{*}a_{1}if_{\pi} (
m_{D}^{2}-m_{K}^{2})
F^{DK}_0 (m_{\pi}^{2}),\label{eq:sd-t}\\
{\cal C}^{\overline K^0 \pi^0} &=&\frac{G_{F}}{\sqrt{2}}V_{ud}V_{cs}^*
a_{2}if_{K} (m_{D}^{2}-m_{\pi}^{2}) F^{D\pi}_0
(m_{K}^{2}),\label{eq:sd-c}
\end{eqnarray}
where the superscripts denote the decay modes. Here, the
nonfactorizable effects, including the radiative corrections to
the weak vertex and the spectator interactions, are absorbed into
the parameters $a_{1,2}$ which amount to replace $N_c$ (= the
number of color) by $N_c^{\rm eff}$ such that
\begin{eqnarray}
a_{1,2}=c_{2,1}+c_{1,2}\frac{1}{N_c^{\rm eff}}\,.
\end{eqnarray}
We have summarized the factorization decay amplitudes in
Appendix~\ref{app:fac}, where the physical $\eta'$ and $\eta$
states are related to the SU(3) octet state $\eta_8$ and singlet
state $\eta_0$ by
\begin{equation}
\left(
\begin{array}{c}
|\eta \rangle\\
      |\eta^\prime \rangle
\end{array}
\right)= \left(
\begin{array}{cc}
\cos\vartheta &-\sin\vartheta\\
\sin\vartheta &\cos\vartheta
\end{array}
\right) \left(
\begin{array}{c}
|\eta_8 \rangle\\
      |\eta_0 \rangle
\end{array}
\right),
\end{equation}
with the mixing angle $\vartheta=-15.4^\circ$
\cite{Feldmann:1998vh}, and
\begin{eqnarray}
|\eta_0\rangle=\frac{1}{\sqrt{3}}|\bar uu+\bar dd+\bar ss\rangle,
\qquad |\eta_8\rangle =\frac{1}{\sqrt{6}}|\bar uu+\bar dd-2\bar
ss\rangle.
\end{eqnarray}
Introducing the decay constants $f_8$ and $f_0$ by
 \begin{eqnarray}
 \langle 0|A_\mu^0|\eta_0\rangle=if_0 p_\mu, \qquad
 \langle 0|A_\mu^8|\eta_8
 \rangle=if_8 p_\mu,
 \end{eqnarray}
we have
 \begin{eqnarray}
f_{\eta'}^u=\frac{f_8}{\sqrt{6}}\sin\vartheta+\frac{f_0}{\sqrt{3}}\cos\vartheta,
\qquad
f_{\eta'}^s=-2\frac{f_8}{\sqrt{6}}\sin\vartheta+\frac{f_0}{\sqrt{3}}\cos
\vartheta,\end{eqnarray} and
\begin{eqnarray}
f_{\eta}^u=\frac{f_8}{\sqrt{6}}\cos\vartheta-\frac{f_0}{\sqrt{3}}\sin\vartheta,
\qquad
f_{\eta}^s=-2\frac{f_8}{\sqrt{6}}\cos\vartheta-\frac{f_0}{\sqrt{3}}\sin
\vartheta, \end{eqnarray}
 where
 \begin{eqnarray}
 \langle 0| \bar u\gamma_\mu\gamma_5 u |\eta^{(\prime)}(p)\rangle = i
 f^u_{\eta^{(\prime)}} p_\mu, \qquad
  \langle 0| \bar s\gamma_\mu\gamma_5 s |\eta^{(\prime)}(p)\rangle = i
 f^s_{\eta^{(\prime)}} p_\mu.
 \end{eqnarray}
The form factors for $B\to \eta^{(\prime)}$ transitions are
assumed to be
 \begin{eqnarray} F^{D\eta}_0 = F^{D\pi}_0
\bigg(\frac{\cos\vartheta}{\sqrt{6}}
-\frac{\sin\vartheta}{\sqrt{3}}\bigg),
 \qquad
  F^{D\eta^{\prime}}_0 =
F^{D\pi}_0 \bigg(\frac{\sin\vartheta}{\sqrt{6}}
+\frac{\cos\vartheta}{\sqrt{3}}\bigg).
 \end{eqnarray}

\subsection{SD weak annihilation amplitudes}\label{subsec:ann}

\begin{figure}[t]
\vspace{0cm} \centerline{\epsfig{file=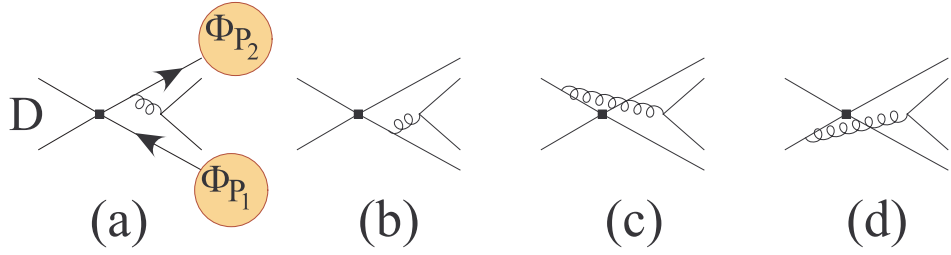,clip=46mm,
width=13cm} } \vspace{0cm} \caption{\small Annihilation
corrections to $D\to P_1\, P_2$, where (a) and (b) correspond to
$A^f_1$, while (c) and (d) give rise to $A^i_1$.} \label{fig:ann}
\end{figure}

The SD weak annihilation
contributions~\cite{Beneke:2001ev,Beneke:2003zv} to $D \to P_1
P_2$, graphically shown in Fig.~\ref{fig:ann}, are represented as
\begin{eqnarray}\label{eq:weakann}
\frac{G_F}{\sqrt2}\!\! \sum_{q,q^{\prime}=d,s}V_{uq}V_{cq^{\prime
}}^{\ast }\! \! \langle P_1 P_2 |\!{\cal T_B} |D\rangle \equiv
A_{\cal T_B} (P_1 P_2) .
\end{eqnarray}
In general, $\langle P_1 P_2 |\!{\cal T_B} |D\rangle$ consists of
$i c f_D f_{P_1} f_{P_2} b_{1,2}$, where $c$ contains factors of
$\pm 1, \pm 1/{\sqrt{2}}$, $1/{\sqrt{6}}$, or $ -2/{\sqrt{6}} $,
arising from the flavor structures of final state mesons, and
\begin{eqnarray}
\label{eq:b1} b_{1,2} &=& \frac{C_F}{N_c^2} c_{1,2}
 A_1^{i}(P_2\, P_1)\,,
 \end{eqnarray}
with the convention adopted here that $P_2$ ($P_1$) contains a
quark (antiquark) arising from the weak vertex with longitudinal
momentum fraction $x$ ($\bar y$). Here the basic building blocks
for annihilation amplitudes originating from operators $(\bar q_1
c)_{V-A} (\bar{q}_2 q_3 )_{V-A} $ are denoted as $A_{1}^{i,f}$,
where the superscript $i \,(f)$ indicates gluon emission from the
initial- (final-) state quarks in the weak vertex, given by
\begin{eqnarray}\label{eq:blocks}
   A_1^i (P_2\, P_1)&=& \pi\alpha_s\! \int_0^1\! dx dy\,
    \left\{ \Phi_{P_2}(x)\,\Phi_{P_1}(y)
    \left[ \frac{1}{y(1-x\bar y)} + \frac{1}{\bar x^2 y} \right]
    + r_\chi^{P_1} r_\chi^{P_2}\,\Phi_{P_2}^p(x)\,\Phi_{P_1}^p(y)\,
     \frac{2}{\bar x y} \right\} ,
    \nonumber\\
   A_1^f (P_2\, P_1)&=& 0 \,.
\end{eqnarray}
with $r_\chi^{P_i}$ being defined as
\begin{equation}\label{rchi}
   r_\chi^{P_i}(\mu) = \frac{2m_{P_i}^2}{m_c(\mu)\,(m_{q_1}(\mu) + m_{q_2}(\mu))}  \,,
\end{equation}
and $m_{q_1,q_2}$  the current quark masses of the meson
constituents in the $\overline{\rm MS}$ scheme. The relevant
two-parton light-cone distribution amplitudes (LCDAs), up to
twist-3, of a light pseudoscalar meson $P$ are defined as
\cite{BraF}
\begin{eqnarray}\label{pidadef}
  \langle P(p)|\bar q_2(z_2)\gamma_\mu\gamma_5 q_1(z_1)|0\rangle
  &=& - i f_P\,p_\mu \int_0^1\! dx\,
   e^{i(x\,p\cdot z_2+\bar x\,p\cdot z_1)}\,\Phi_P(x) \,, \nonumber\\
  \langle P(p)|\bar q_2(z_2) i\gamma_5 q_1(z_1)|0\rangle
  &=& f_P\mu_P \int_0^1\! dx\,
   e^{i(x\,p\cdot z_2+\bar x\,p\cdot z_1)}\,\Phi_P^p(x) \,, \nonumber\\
  \langle P(p)|\bar q_2(z_2)\sigma_{\mu\nu}\gamma_5 q_1(z_1)|0\rangle
  &=& i f_P\mu_P\,(p_\mu z_\nu - p_\nu z_\mu)
   \int_0^1\! dx\,e^{i(x\,p\cdot z_2+\bar x\,p\cdot z_1)}\,
   \frac{\Phi_P^\sigma(x)}{6} \,,
\end{eqnarray}
where $z=z_2-z_1$, $\mu_P=m_P^2/(m_{q_1}+m_{q_2})$, $f_P$ is the
decay constant, and $x\ ({\rm or}\ \bar x=1-x)$ is the collinear
momentum fraction carried by the quark $q_2$ (or antiquark $\bar
q_1$). Here and below we do not explicitly show the gauge factors
\begin{equation}
{\rm P}\exp\left[ig_s\!\!\int_0^1\!\! dt\,(z_1-z_2)_\mu
  A^\mu(t z_1+(1-t)z_2)\right]
\label{Pexp}
\end{equation}
in between the quark fields. The leading-twist LCDA $\Phi_P(x)$ is
of twist-2, while $\Phi_P^p(x)$ and $\Phi_P^\sigma(x)$ are of
twist-3. LCDAs appearing in the calculation of weak annihilation
contributions are in the form of
\begin{eqnarray}\label{eq:spacedef}
   &&\langle P(p)|\bar q_{2,\beta}(z_2)\,q_{1,\alpha}(z_1)|0\rangle
    \nonumber\\
   &&= \frac{i f_P}{4} \int_0^1\! dx\,
    e^{i(x\,p\cdot z_2+\bar x\,p\cdot z_1)}
    \Bigg\{ \slash\!\!\! p\,\gamma_5\,\Phi_P(x)
    - \mu_P\gamma_5 \Bigg( \Phi_P^p(x) - \sigma_{\mu\nu}\,p^\mu z^\nu\,
    \frac{\Phi_P^\sigma(x)}{6} \Bigg) \Bigg\}_{\alpha\beta} . \quad
\end{eqnarray}
Neglecting three-particle contributions, the twist-3 distribution
amplitudes in the asymptotic limit are related to each other by
equations of motion, so that
\begin{equation}
   \Phi_P^p(x) = 1 \,, \qquad
   \frac{\Phi^{\sigma\,\prime}_P(x)}{6} = (\bar x-x)\Phi_P^p \,, \qquad
   \frac{\Phi_P^\sigma(x)}{6} = (x\bar x)\Phi_P^{p} \,.
\end{equation}
Using the above simplification, one can get the corresponding
projector of Eq.~(\ref{eq:spacedef}) in the momentum
space~\cite{Geshkenbein:qn,Beneke:2001ev,Beneke:2003zv}
\begin{equation}\label{eq:piprojector}
   M_{\alpha\beta}^P = \frac{i f_P}{4} \left(
   \not p\,\gamma_5\,\Phi_P(x) - \mu_P\gamma_5\,
   \frac{\not k_2\,\not k_1}{k_2\cdot k_1}\,\Phi_P^p(x)
   \right)_{\alpha\beta},
\end{equation}
and further obtain the basic building blocks for annihilation
amplitudes given in Eq.~(\ref{eq:blocks}), where the momenta of
the quark $q_1$ and anitiquark $\bar q_2$ in a meson are
parameterized as
\begin{eqnarray}
k_1^\mu = x E n_-^\mu +k_\perp^\mu + \frac{k_\perp^2}{4
xE}n_+^\mu\,, \ \ \ \ k_2^\mu = \bar x E n_-^\mu - k_\perp^\mu +
\frac{k_\perp^2}{4 \bar x E}n_+^\mu,
\end{eqnarray}
respectively. For simplicity, we have introduced two light-like
vectors $n_-^\mu\equiv (1,0,0,-1), n_+^\mu\equiv (1,0,0,1)$. If
neglecting the meson mass squared, we have $p^{\mu} = E n_-^\mu$
where $E$ is the energy of the meson. We refer the reader to
Refs.~\cite{Beneke:2001ev,Beneke:2003zv} for the detailed
technique of calculating weak annihilation contributions.

The LCDAs normalized at the scale $\mu$ can be expanded in
Gegenbauer polynomials of forms
\begin{eqnarray}
   \Phi_{P}(x,\mu) &=& 6x(1-x)\,\bigg[ 1 + \sum_{n=1}^\infty
   a_n^{P}(\mu)\,C_n^{(3/2)}(2x-1) \bigg],\label{gegenbauer1}\\
   \Phi_{P}^p(x,\mu) &=& \,1 + \sum_{n=1}^\infty
   a_n^{P,p}(\mu)\,C_n^{(1/2)}(2x-1),\label{gegenbauer2} \\
    \Phi_{P}^{\sigma} (x,\mu) &=& 6x(1-x)\,\bigg[ 1 + \sum_{n=1}^\infty
   a_n^{P,\sigma}(\mu)\,C_n^{(3/2)}(2x-1) \bigg]\label{gegenbauer3}.
\end{eqnarray}
In the numerical analysis, we truncate the expansion of $\Phi_P$
at $n=1$ and just take the asymptotic approximation for $\Phi_P^p$
and $\Phi_P^\sigma$.
Note that $a_1^P$ is nonzero only for the kaon. For the kaon
containing an $\bar s$ quark, we have the replacement $x
\leftrightarrow \bar x$ in Eq.~(\ref{gegenbauer1}). The
annihilation corrections to $D^0 \to K^0 \overline K^0$, as an
example, thus read
\begin{eqnarray}\label{eq:XAK0K0}
&& A_{\cal T_B} (K^0 \overline K^0)=
  i \frac{G_F}{\sqrt2} f_D f_{K}^2
  \frac{C_F}{N_c^2} \pi\alpha_s  c_{1} \nonumber\\
 &&\ \ \  \times \Bigg\{ V_{us} V_{cs}^* \int_0^1\! dx dy\,
    \Bigg[ \Phi_{\overline K^0}(x)\,\Phi_{K^0}(y)
    \left( \frac{1}{y(1-x\bar y)} + \frac{1}{\bar x^2 y}
    \right)
  + (r_{\chi}^K)^2\,\Phi_{\overline K^0}^p(x)\,\Phi_{K^0}^p(y)\,
     \frac{2}{\bar x y} \Bigg]\nonumber\\
 && \ \ \ \ +  V_{ud} V_{cd}^* \int_0^1\! dx dy\,
    \Bigg[ \Phi_{K^0}(x)\,\Phi_{\overline K^0}(y)
    \left( \frac{1}{y(1-x\bar y)} + \frac{1}{\bar x^2 y}
    \right)
 + (r_{\chi}^K)^2\,\Phi_{K^0}^p(x)\,\Phi_{\overline K^0}^p(y)\,
     \frac{2}{\bar x y} \Bigg] \Bigg\}\nonumber\\
 && =i \frac{G_F}{\sqrt2} f_D f_{K}^2
  \frac{C_F}{N_c^2} \pi\alpha_s  c_{1}  V_{us} V_{cs}^* 36 a_1^K
  (4X_A+33-4\pi^2),
\end{eqnarray}
where use of $V_{ud} V_{cd}^*=-V_{us} V_{cs}^*$ has been made, and
$\int_0^1 dz/z \rightarrow X_A$ has been used to parameterize the
logarithmically divergent
integrals~\cite{Beneke:2001ev,Beneke:2003zv}, which can be
regulated by including the transverse momentum of the quark in the
end point region of integrals, but however may suffer from some
theoretical problems (see discussions in the introduction). It is
interesting to note that $A_{\cal T_B} (K^0 \overline K^0)$ is
proportional to $a_1^K$. As will be seen in
Sec.~\ref{sec:results}, the magnitude of $a_1^{K}$ has a large
impact on the $D^0 \to K^0 \overline K^0$ branching ratio. Two
remarks are in order. First, the simplified form of the projector
in Eq.~(\ref{eq:piprojector}) cannot be justified if considering
higher Gegenbauer moment corrections to $\Phi_P^p$ and
$\Phi_P^\sigma$. We have checked that the amplitude corrections due to
$a_1^{K,p}$ and $a_1^{K,\sigma}$ are numerically negligible if the
magnitudes of $a_1^{K,p}$ and $a_1^{K,\sigma}$ are not too large.
Second, we do not consider $a_2^P$, since distinguishing $a_2^\pi,
a_2^K$, and $a_2^{\eta_8}$ is not numerically significant in the
present study~\cite{Beneke:2001ev}, and, moreover, partial effects
due to $a_2^P$ can be absorbed in $X_A$. The detailed expressions
for SD weak annihilation amplitudes are collected in
Appendix~\ref{app:ann}.

\section{SU(3) rescatterings}\label{sec:fsi}

From the isospin amplitude analysis of $D^{0}\rightarrow K^{-}\pi
^{+}, \overline{K}^{0}\pi^{0}$ and $D^{+}\rightarrow
\overline{K}^{0}\pi^{+}$, as discussed in
Sec.~\ref{sec:introduction}, we know that the LD FSIs effects may
be significant in $D$ meson decays. Considering elastic SU(3)
rescatterings in $D$ decays, for instance, $D^{0}\rightarrow
K^{-}\pi^{+}$, $D^{0}\rightarrow \overline{K}^{0}\pi^{0}$ and
$D^{0}\rightarrow \overline{K}^{0}\eta_{8}$ can be generated via
\begin{equation*}
\begin{array}{ccc}
& \rightarrow & K^{-}\pi^{+} \\
D^{0}\rightarrow K^{-}\pi^{+} & \rightarrow & \overline{K}^{0}\pi^{0} \\
& \rightarrow & \overline{K}^{0}\eta_{8}
\end{array}
,\hspace{0.5cm}
\begin{array}{ccc}
& \rightarrow & K^{-}\pi^{+} \\
D^{0}\rightarrow \overline{K}^{0}\pi^{0} & \rightarrow &
\overline{K}^{0}\pi^{0} \\
& \rightarrow & \overline{K}^{0}\eta_{8}
\end{array}
,\hspace{0.5cm}
\begin{array}{ccc}
& \rightarrow & K^{-}\pi^{+} \\
D^{0}\rightarrow \overline{K}^{0}\eta_{8} & \rightarrow & \overline{K}^{0}\pi^{0} \\
& \rightarrow & \overline{K}^{0}\eta_{8}
\end{array}
.
\end{equation*}
Taking into account elastic SU(3) FSIs, the decay amplitudes
$\mathbf{A}_{i}^{\rm FSI}$ are given
by~\cite{Watson,Suzuki:1999uc,Smith:1998nu}
\begin{eqnarray}
{\mathbf{A}_{i}^{\rm FSI}}&=& \sum_l {\mathbf S}^{1/2}_{il}
{\mathbf A}_{l}^{\rm bare} = (\mathbf{U}^{\rm T}{\mathbf
S^{1/2}}_{\rm diag}\,\mathbf{U})_{il} {\mathbf A}_{l}^{\rm bare},
\label{eq:fsi}
\end{eqnarray}
where $\mathbf{S}$ is strong interaction scattering matrix, and
$\mathbf{A}_{l}^{\rm bare} (=\mathbf{A}_{l}^{\rm
fac}+\mathbf{A}_{l}^{\cal T_B})$ are approximated in terms of the
factorization and SD weak annihilation amplitudes. Note that
$\mathbf{S}$ is unitary. The SU(3) final-state rescatterings for
$D\to P_1 P_2$ are described by the product $\mathbf{8\otimes 8}$.
Since the $P_1 P_2$ states obey the Bose symmetry, only the
symmetric states given by the representation $\mathbf{36(=27\oplus
8\oplus 1)}$ in $\mathbf {8 \otimes 8(= 36 \oplus 28)}$
decomposition are relevant, whereas states given by the
representation $\mathbf{28(=10\oplus \overline {10}\oplus 8)}$
vanish.

In the present study, we will use $\delta_{27}$, $\delta_{8}$ and
$\delta_{1}$ to stand for the respective rescattering phases of
$\mathbf{27}$, $\mathbf{8}$ and $\mathbf{1}$ states. The detailed
derivation for $\mathbf{U}$ matrices and the corresponding SU(3)
eigen-amplitudes is exhibited in Appendix~\ref{appsec:fsi}. Thus
the $\mathbf{S}^{1/2}$ matrices and decay amplitudes can be recast
into the following 5 subsets (see also
Refs.~\cite{Chua:2002wk,Smith:1998nu}):
\begin{itemize}
 \item subset 1
($K^{-}\pi^{+}-\overline{K}^{0}\pi^{0}-\overline{K}^{0}\eta_{8}$
rescatterings),
\begin{equation}\label{eq:subset1.1}
\mathbf{S}_{(\overline{K\pi})^{0}}^{1/2}e^{-i\delta_{27}}
 =\left(
\begin{array}{ccc}
\frac{2+3e^{-i\delta }}{5} & \frac{3\left(
1-e^{-i\delta}\right)}{5\sqrt{2}}
 & \frac{\sqrt{3}\left( 1-e^{-i\delta }\right) }{5\sqrt{2}} \\
\frac{3\left( 1-e^{-i\delta }\right) }{5\sqrt{2}} &
\frac{7+3e^{-i\delta }}{10}
 & \frac{\sqrt{3}\left( -1+e^{-i\delta }\right) }{10} \\
\frac{\sqrt{3}\left( 1-e^{-i\delta }\right) }{5\sqrt{2}} &
\frac{\sqrt{3} \left( -1+e^{-i\delta }\right) }{10} &
\frac{9+e^{-i\delta }}{10}
\end{array}
\right) ,
\end{equation}
\begin{equation}\label{eq:subset1.2}
\mathbf{A}^{\rm bare}_{( \overline{K\pi} )^{0}}=\left(
\begin{array}{c}
A_{K^{-}\pi^{+}}^{\rm bare} \\
A_{\overline{K}^{0}\pi^{0}}^{\rm bare} \\
A_{\overline{K}^{0}\eta_{8}}^{\rm bare}
\end{array}
\right) ,
\end{equation}
\item subset~2 ($ K^{+}\pi^{-}-K^{0}\pi^{0}-K^{0}\eta_{8}$\
rescatterings),
\begin{equation}\label{eq:subset2.1}
\mathbf{S}_{({K\pi})^{0}}^{1/2}=\mathbf{S}_{(\overline{K\pi})^{0}}^{1/2},
\end{equation}
\begin{equation}\label{eq:subset2.2}
\mathbf{A}_{( {K\pi} )^{0}}^{\rm bare}=\left(
\begin{array}{c}
A_{K^{+}\pi^{-}}^{\rm bare} \\
A_{K^{0}\pi^{0}}^{\rm bare} \\
A_{K^{0}\eta_{8}}^{\rm bare}
\end{array}
\right) ,
\end{equation}
\item subset~3 ($K^{0}\pi^{+}-K^{+}\pi^{0}-K^{+}\eta_{8}$
rescatterings),
\begin{equation}\label{eq:subset3.1}
\mathbf{S}_{\left( K\pi \right)^{+}}^{1/2}e^{-i\delta
_{27}}=\left(
\begin{array}{ccc}
\frac{2+3e^{-i\delta }}{5} & -\frac{3\left( 1-e^{-i\delta }\right) }{5\sqrt{2%
}} & -\frac{\sqrt{3}\left( 1-e^{-i\delta }\right) }{5\sqrt{2}} \\
-\frac{3\left( 1-e^{-i\delta }\right) }{5\sqrt{2}} & \frac{7+3e^{-i\delta }}{%
10} & -\frac{\sqrt{3}\left( -1+e^{-i\delta }\right) }{10} \\
-\frac{\sqrt{3}\left( 1-e^{-i\delta }\right) }{5\sqrt{2}} & -\frac{\sqrt{3}%
\left( -1+e^{-i\delta }\right) }{10} & \frac{9+e^{-i\delta }}{10}
\end{array}
\right) ,
\end{equation}
\begin{equation}\label{eq:subset3.2}
\mathbf{A}_{( K\pi )^{+}}^{\rm bare}=\left(
\begin{array}{c}
A_{K^{0}\pi^{+}}^{\rm bare} \\
A_{K^{+}\pi^{0}}^{\rm bare} \\
A_{K^{+}\eta_{8}}^{\rm bare}
\end{array}
\right) ,
\end{equation}
\item subset~4
($\pi^{+}\pi^{0}-\pi^{+}\eta_{8}-K^{+}\overline{K}^{0}$
rescatterings),
\begin{equation}\label{eq:subset4.1}
\mathbf{S}_{\left( \pi \pi
\right)^{+}}^{1/2}e^{-i\delta_{27}}=\left(
\begin{array}{ccc}
1 & 0 & 0 \\
0 & \frac{3+2e^{-i\delta }}{5} & -\frac{\sqrt{6}\left(
1-e^{-i\delta
}\right) }{5} \\
0 & -\frac{\sqrt{6}\left( 1-e^{-i\delta }\right) }{5} &
\frac{2+3e^{-i\delta }}{5}
\end{array}
\right) ,
\end{equation}
\begin{equation}\label{eq:subset4.2}
\mathbf{A}_{( \pi \pi)^{+}}^{\rm bare}=\left(
\begin{array}{c}
A_{\pi^{+}\pi^{0}}^{\rm bare} \\
A_{\pi^{+}\eta_{8}}^{\rm bare} \\
A_{K^{+}\overline{K}^{0}}^{\rm bare}
\end{array}
\right) ,
\end{equation}
\item subset 5
($\pi^{+}\pi^{-}-\pi^{0}\pi^{0}-\eta_{8}\eta_{8}-K^{+}K^{-}-
K^{0}\overline{K}^{0}-\pi^{0}\eta_{8}$ rescatterings),
\begin{eqnarray}\label{eq:subset5.1}
&&{\hspace{-1.1cm}}  \mathbf{S}_{\left( \pi \pi \right)^{0}}^{1/2}e^{-i\delta_{27}} \\
&&{\hspace{-1.1cm}} \left(
\begin{array}{cccccc}
\frac{5e^{-i\sigma }+8e^{-i\delta }+7}{20} & \frac{5e^{-i\sigma
}+8e^{-i\delta }-13}{20\sqrt{2}} & \frac{5e^{-i\sigma
}-8e^{-i\delta }+3}{20 \sqrt{2}} & \frac{5e^{-i\sigma
}-4e^{-i\delta }-1}{20} & \frac{5e^{-i\sigma
}-4e^{-i\delta }-1}{20} & 0 \\
\frac{5e^{-i\sigma }+8e^{-i\delta }-13}{20\sqrt{2}} &
\frac{5e^{-i\sigma }+8e^{-i\delta }+27}{40} & \frac{5e^{-i\sigma
}-8e^{-i\delta }+3}{20\sqrt{2}} & \frac{5e^{-i\sigma
}-4e^{-i\delta }-1}{20\sqrt{2}} & \frac{5e^{-i\sigma
}-4e^{-i\delta }-1}{20\sqrt{2}} & 0 \\
\frac{5e^{-i\sigma }-8e^{-i\delta }+3}{20\sqrt{2}} &
\frac{5e^{-i\sigma }-8e^{-i\delta }+3}{20\sqrt{2}} &
\frac{5e^{-i\sigma }+8e^{-i\delta }+27}{40} & \frac{5e^{-i\sigma
}+4e^{-i\delta }-9}{20\sqrt{2}} & \frac{5e^{-i\sigma
}+4e^{-i\delta }-9}{20\sqrt{2}} & 0 \\
\frac{5e^{-i\sigma }-4e^{-i\delta }-1}{20} & \frac{5e^{-i\sigma
}-4e^{-i\delta }-1}{20\sqrt{2}} & \frac{5e^{-i\sigma
}+4e^{-i\delta }-9}{20 \sqrt{2}} & \frac{5e^{-i\sigma
}+8e^{-i\delta }+7}{20} & \frac{5e^{-i\sigma }-4e^{-i\delta
}-1}{20} & \frac{4\sqrt{3}\left( e^{-i\delta }-1\right) }{20}
\\
\frac{5e^{-i\sigma }-4e^{-i\delta }-1}{20} & \frac{5e^{-i\sigma
}-4e^{-i\delta }-1}{20\sqrt{2}} & \frac{5e^{-i\sigma }+4e^{-i\delta }-9}{20%
\sqrt{2}} & \frac{5e^{-i\sigma }-4e^{-i\delta }-1}{20} &
\frac{5e^{-i\sigma }+8e^{-i\delta }+7}{20} & \frac{4\sqrt{3}\left(
1-e^{-i\delta }\right) }{20}
\\
0 & 0 & 0 & \frac{4\sqrt{3}\left( e^{-i\delta }-1\right) }{20} & \frac{4%
\sqrt{3}\left( 1-e^{-i\delta }\right) }{20} & \frac{4\left(
2e^{-i\delta }+3\right) }{20}
\end{array}
\right) .  \nonumber
\end{eqnarray}
\begin{equation}\label{eq:subset5.2}
\mathbf{A}_{(\pi \pi)^{0}}^{\rm bare}=\left(
\begin{array}{c}
A_{\pi^{+}\pi^{-}}^{\rm bare} \\
A_{\pi^{0}\pi^{0}}^{\rm bare} \\
A_{\eta_{8}\eta_{8}}^{\rm bare} \\
A_{K^{+}K^{-}}^{\rm bare} \\
A_{K^{0}\overline{K}^{0}}^{\rm bare} \\
A_{\pi^{0}\eta_{8}}^{\rm bare}
\end{array}
\right) ,
\end{equation}
\end{itemize}
where $\delta \equiv \delta_{27}-\delta_{8}$, $\sigma \equiv
\delta_{27}-\delta_{1}$, and we have included the identical
particle factor $1/\sqrt2$ in the amplitudes ${\cal
A}_{\pi^0\pi^0}^{\rm bare}$ and ${\cal A}_{\eta_8\eta_8}^{\rm
bare}$. Here $\mathbf{S}^{1/2}$ matrices have been factored out an
overall phase $e^{i\delta_{27}}$ since only phase differences
affect physical results. Note that we do not list $D^+ \to
\overline K^0 \pi^+$, which does not belong to any above subset,
i.e., does not rescatter with other $PP$ modes. Note also that in
the subset~4, $D^+ \to \pi^+ \pi^0$ does not rescatter with other
modes, too.

\section{Results}\label{sec:results}

In this section, we will first introduce the relevant parameters
in the fit and then give the numerical results together with a
brief discussion.
The  2-body $D$ meson decay rates are given by
\begin{equation}
\Gamma \left( D\rightarrow P_{1}P_{2}\right)
=\frac{|\overrightarrow{p_c}| }{8\pi m_D^{2}} |A^{\rm FSI}|^{2},
\end{equation}
where $\overrightarrow{p_c}$ is the center-of-mass momentum of
decay particles. In the numerical analysis, we perform the best
multi-mode $\chi^2$ fit for measured branching ratios, defined as
\begin{eqnarray}
\chi^2=\sum_i \Big(\frac{y_i-x_i}{\Delta x_i}\Big)^2,
\label{chi2}\end{eqnarray} where $y_i$ and $x_i\pm \Delta x_i$
denote the theoretical results and measurements, respectively. On
the theoretical side, input parameters relevant for our numerical
analysis are listed in
Table~\ref{tab:inputs}~\cite{PDG,Semenov:2003ne,Chen:1999nx,Buchalla:1995vs}.
As listed in Table~\ref{tab:results}, we take the current
data~\cite{PDG} for the 14 $K \pi$, $\pi\pi$, $KK$,
$K\eta^{(\prime)}$ and $\pi\eta^{(\prime)}$ BRs as inputs. The
modes involving $\eta$ or $\eta^\prime$ are related to $\eta_8$
and $\eta_0$ via the mixing angle $\vartheta$. The SU(3) FSI
picture is not suitable to be extended to the U(3) scenario since
U$_A$(1) symmetry is broken by anomaly, i.e., $\eta'$ is not a
Goldstone boson. The weak annihilation effect for SU(3) channels
is parameterized in terms of $X_A$, while that for decay modes
involving $\eta_0$ is distinguished to be $X_A'$. However we do
not distinguish $1/N_c^{\rm eff}$ because it is numerically small,
as seen in our analysis. The scale for the factorization
amplitudes is taken to be $\mu=m_c$, i.e., $1/N_c^{\rm
eff}=1/N_c^{\rm eff}(m_c)$, while the scale for SD weak
annihilation amplitudes is 1~GeV. We use the world average value
of $F_0^{DK}(0)=0.76\pm 0.03$~\cite{Semenov:2003ne}. For the $q^2$
dependence of form factors, we adopt the pole dominance
assumption:
\begin{eqnarray}
F_0(q^2) = \frac{F_0(0)}{1-q^2/m^2_*},
\end{eqnarray}
with taking $m_*$ as the mass of the lowest-lying scalar charmed
meson in the corresponding channel. The above form is consistent
with the recent QCD sum rule study for $B\to light\ meson$
transitions~\cite{Ball:2004hn}. We assume
$m_*=2.3$~GeV~\cite{Colangelo:2003vg} (or $2.2$~GeV) for
$F_0^{DK}$ (or $F_0^{D\pi}$). The results for fitted parameters,
which are (i) two FSI phases, $\delta$ and $\sigma$, (ii) the form
factor $F_0^{D\pi}$, (iii) SD weak annihilation parameter $X_A$
and $X_A'$, and (iv) $1/N_c^{\rm eff}$, are cataloged in
Table~\ref{tab:phase}. $\it Output \ observables$ are given in
Table~\ref{tab:results}. The errors of outputs correspond to the
variation of $F_0^{DK}(0)$, while the errors due to uncertainties
of $D$ lifetimes are negligible.

The nonfactorizable effects are lumped into the effective number
of color $N_c^{\rm eff}$, of which the deviation from $N_c$
measures such effects. $1/N_c^{\rm eff}$ could be complex. However
it is assumed to be real due to its small value: $1/N_c^{\rm eff}<
-1/15 \ (\simeq -0.067)$ in the fit, consistent with the very
earlier large-$N_c$ approach for describing hadronic $D$
decays~\cite{Buras:1985xv}. It is interesting to note that we
obtain the weak annihilation parameter $|X_A| = 3.84\pm 0.06$
($|X_A'| = 2.45^{+0.07}_{-0.46}$ or $2.18\pm 0.19$)  with a large
phase $(-138\pm 3)^\circ$ ($(-138\pm 3)^\circ$ or $(130\pm
3)^\circ$), compared with the similar parameter $|X_A|\sim 4.5$
given in $B$ decays~\cite{Beneke:2003zv,Kagan:2004uw,Yang:2005tv}.
Note that we obtain a twofold solution for $X_A'$. As seen in
Table~\ref{tab:results}, the weak annihilation topologies have a
large impact on branching ratios. This analysis gives moderate
rescattering phases $\delta\simeq -46^\circ$ and $\sigma\simeq
-21^\circ$.
\begin{table}[t]
\centerline{\parbox{14cm}{\caption{\label{tab:inputs} Summary of
input
parameters~\cite{PDG,Semenov:2003ne,Chen:1999nx,Buchalla:1995vs}
on the theoretical side of the fit.}}} \vspace{0.1cm}
\begin{center}
{\tabcolsep=0.699cm\begin{tabular}{|c|c|c|c|c|} \hline\hline
\multicolumn{5}{|c|}{Running quark masses [GeV] and the strong coupling constant} \\
\hline   $m_c(m_c)$ & $m_s(1\,\mbox{GeV})$
       & $m_{u}(1\,\mbox{GeV})$ & $m_d(1\,\mbox{GeV})$ & $\alpha_s(1~{\rm GeV})$ \\
\hline
 $1.35$ & $0.12$ & $0.004$ & $0.009$ & $0.517$\\
\hline
\end{tabular}}
{\tabcolsep=1.87cm\begin{tabular}{|c|c|c|} \hline
\multicolumn{3}{|c|}{The Wolfenstein parameter
 and $D$-meson lifetimes [$10^{-15}s$]} \\
\hline  $\lambda$ &  $\tau(D^+)$ & $\tau(D^0)$ \\
\hline  $0.2196$ & $1040\pm 7$ & $410.3\pm 1.5$   \\
\hline
\end{tabular}}
{\tabcolsep=1.18cm\begin{tabular}{|c|c|c|c|c|} \hline
\multicolumn{5}{|c|}{Pseudoscalar-meson decay constants [MeV]} \\
\hline
$f_\pi$ & $f_K$ & $f_{\eta_8}$ & $f_{\eta_0}$ & $f_D$ \\
\hline
131 & 160 & $168$ & $157$ & $220\pm 20$\\
\hline
\end{tabular}}
{\tabcolsep=3.145cm\begin{tabular}{|c|c|} \hline
\multicolumn{2}{|c|}{The form factor (at $q^2=0$) and $\eta-\eta'$ mixing angle} \\
\hline $F_0^{DK}(0)$ & $\vartheta$\\
\hline
$0.76\pm 0.03$ & $-14.5^\circ$ \\
\hline
\end{tabular}}
{\tabcolsep=1.226cm\begin{tabular}{|c|c|c|c|} \hline
\multicolumn{4}{|c|}{The Wilson coefficients for $D$ decays} \\
\hline
$c_1(m_c)$ & $c_2(m_c)$ & $c_1(1~{\rm GeV})$ & $c_2(1~{\rm GeV}$) \\
\hline
$1.216$ & $-0.422$ & $1.275$ & $-0.510$\\
\hline
\end{tabular}}
\end{center}
\end{table}

\begin{table*}[t!]
\vspace{1.5cm} \caption{ \label{tab:results}\small The branching
ratios in units of $10^{-3}$: data (${\cal
B}_{\text{Exp}}$)~\cite{PDG} vs. fitted results (${\cal
B}_{\text{FSI}}$). The individual $\chi^{2}_i$ values of decay
modes corresponding to the best fit are listed. For comparison,
taking the best fit parameters and $F_0^{DK}(0)=0.76$ into
account, we then give (i) ${\cal B}_{\text{Fact}}$ by means of
setting $\delta=\sigma=0$ and neglecting the weak annihilation
corrections, (ii) ${\cal B}_{\text{NoAnn }}$ by means of
neglecting only the SD weak annihilation corrections, and (iii)
${\cal B}_{\text{NoFSI}}$  by means of setting $\delta=\sigma=0$.
Note that $D^0\to \eta'\eta'$ is kinematically forbidden. The
errors in ${\cal B}_{\text{FSI}}$ and $\chi^{2}_i$ are due to the
variation of $F_0^{DK}(0)$.}
%
%
\begin{center}
{\tabcolsep=0.3cm\begin{tabular}{|c|c|ccc|cc|} \hline\hline
 Decay modes & ${\cal B}_{\text{Exp}}$ & ${\cal B}_{\text{Fact}}$
 & ${\cal B}_{\text{NoAnn}}$ & ${\cal B}_{\text{NoFSI}}$
 & ${\cal B}_{\text{FSI}}$ & $\chi^{2}_i$ \\
\hline $D^{0}\rightarrow K^{-}\pi^{+}$ & $38.0\pm 0.9$ &
 $62.20$ & $56.20$ & $56.69$ & $38.02^{+0.05}_{-0.09}$ & $0.00^{+0.04}_{-0.00}$ \\
$D^{0}\rightarrow \overline{K}^{0}\pi^{0}$ & $23.0\pm 2.2$
 & $10.39$ & $14.05$ & $16.69$ & $23.83^{+0.01}_{-0.19}$ & $0.14^{+0.01}_{-0.06}$ \\
$D^{0}\rightarrow \overline{K}^{0}\eta$ & $7.7\pm 1.1$
 & $2.75$ & $3.83$ & $ 2.67$ & $7.92^{+0.17}_{-0.05}$ & $0.04^{+0.09}_{-0.02}$ \\
$D^{0}\rightarrow \overline{K}^{0}\eta^{\prime}$
 & $18.8\pm 2.8$ & $2.40$ & $3.14$ & $ 16.11$ & $19.82^{+0.17}_{-0.05}$ & $0.13^{+0.05}_{-0.01}$\\
\hline $D^{+}\rightarrow K^{0}\pi^{+}$ & ---
 & $0.14$ & $0.21$ & $0.09$ & $0.27^{+0.01}_{-0.02}$ & --- \\
$D^{+}\rightarrow K^{+}\pi^{0}$ & ---
 & $0.46$ & $0.39$ & $0.59$ & $0.39\pm 0.05$ & --- \\
$D^{+}\rightarrow K^{+}\eta$ & ---
 & $0.12$ & $0.10$ & $0.08$ & $ 0.07\pm 0.00$ & --- \\
$D^{+}\rightarrow K^{+}\eta^{\prime}$ & ---
 & $0.11$ & $0.12$ & $0.19$ & $ 0.20\pm 0.01$ & --- \\
\hline $D^{+}\rightarrow \overline{K}^{0}\pi^{+}$ & $28.2\pm 1.9$
& $28.15$ & $ 28.15$ & $28.15$ & $28.15^{+0.05}_{-0.03}$ & $0.00\pm 0.00$ \\
\hline $D^{0}\rightarrow K^{+}\pi^{-}$ & $0.138\pm 0.011$
 & $0.36$ & $0.31$ & $ 0.24$ & $0.15\pm 0.00$ & $0.68^{+0.28}_{-0.16}$ \\
$D^{0}\rightarrow K^{0}\pi^{0}$ & ---
 & $0.03$ & $0.06$ & $0.03$ & $ 0.08\pm 0.00$ & --- \\
$D^{0}\rightarrow K^{0}\eta$ & ---
 & $0.007$ & $0.02$ & $0.006$ & $ 0.03\pm 0.01$ & --- \\
$D^{0}\rightarrow K^{0}\eta^{\prime}$ & ---
 & $0.006$ & $0.009$ &$0.06$ & $0.07\pm 0.00$ & --- \\
\hline $D^{+}\rightarrow \pi^{+}\pi^{0}$ & $2.6\pm 0.7$
 & $2.27$ & $2.27$ & $ 2.27$ & $2.27^{+0.03}_{-0.04}$ & $0.22^{+0.05}_{-0.04}$ \\
\hline $D^{+}\rightarrow K^{+}\overline{K}^{0}$ & $5.9\pm 0.6$
 & $11.67$ & $10.47$ & $6.77$ & $5.73^{+0.29}_{-0.23}$ & $0.08^{+0.36}_{-0.04}$ \\
$D^{+}\rightarrow \pi^{+}\eta$ & $3.0\pm 0.6$
 & $0.77$ & $2.61$ & $0.45$ & $2.61\pm 0.01$ & $0.42^{+0.02}_{-0.01}$ \\
$D^{+}\rightarrow \pi^{+}\eta^{\prime}$ & $5.1\pm 1.0$
 & $3.47$ & $3.02$ & $4.15$ & $3.31^{+0.14}_{-0.16}$ & $3.19^{+0.58}_{-0.47}$  \\
\hline $D^{0}\rightarrow \pi^{+}\pi^{-}$ & $1.38\pm 0.05$
 & $4.73$ & $4.14$ & $2.76$ & $1.37\pm 0.01$ & $0.02^{+0.01}_{-0.00}$ \\
$D^{0}\rightarrow \pi^{0}\pi^{0}$ & $0.84\pm 0.22$
 & $0.36$ & $0.71$ & $0.22$ & $0.73^{+0.03}_{-0.04}$ & $0.26^{+0.18}_{-0.11}$ \\
$D^{0}\rightarrow K^{+}K^{-}$ & $3.89\pm 0.14$
 & $4.58$ & $3.92$ & $5.43$ & $3.85^{+0.01}_{-0.00}$ & $0.08^{+0.00}_{-0.04}$ \\
$D^{0}\rightarrow K^{0}\overline{K}^{0}$ & $0.71\pm 0.19$
 & $0$ & $0.00$ & $0.65$ & $0.68^{+0.04}_{-0.06}$ & $0.03^{+0.23}_{-0.03}$ \\
$D^{0}\rightarrow \pi^{0}\eta$ & ---
 & $0.09$ & $0.35$ & $0.25$ & $0.68\pm 0.01$ & --- \\
$D^{0}\rightarrow \pi^{0}\eta^{\prime}$ & ---
 & $0.09$ & $0.15$ & $0.01$ & $0.05^{+0.03}_{-0.02}$ & --- \\
$D^{0}\rightarrow \eta\eta$ & ---
 & $0.10$ & $0.44$ & $0.33$ & $1.17^{+0.00}_{-0.02}$ & --- \\
$D^{0}\rightarrow \eta\eta^{\prime}$ & ---
 & $0.13$ & $0.25$ & $1.29$ & $1.95\pm 0.05$ & --- \\
$D^{0}\rightarrow \eta^{\prime}\eta^{\prime}$ & ---
 & $0$ & $0$ & $0$ & $0$ & --- \\
\hline\hline
\end{tabular}}
\end{center}
\vspace{0.5cm}
\end{table*}

\begin{table*}[t!]
\caption{ The $\chi^2_{\rm min}/{\rm d.o.f.}$ and fitted
parameters, where we obtain a twofold solution for $X_A'$ which is
relevant only for decay modes involving $\eta_0$. The errors are
due to the variation of $F_0^{DK}(0)$.
 \label{tab:phase} }
\begin{ruledtabular}
\begin{tabular}{cc}
      &Best fit results
      \\
\hline
 $\chi^2_{\rm min}/{\rm d.o.f.}$
        & $(5.3^{+1.3}_{-0.5})/5$
       \\
 $\delta$
        &  $(-46\pm 2)^\circ$
        \\
 $\sigma$
        & $(-21\pm 1)^\circ$
        \\
 $N_c^{\rm eff}$
       & $-21^{+6}_{-18} $
       \\
 $F_0^{D\pi}(0)$
        & $0.83\pm 0.02$
       \\
 $a_1^K$ & $-0.15^{+0.00}_{-0.01}$
       \\
 $|X_A|$
        & $3.84\pm 0.06$
        \\
 ${\rm arg}(X_A)$
        & $(-138\pm 3)^\circ$
        \\
\hline
 $|X_A'|$ & $2.45^{+0.07}_{-0.46}$ \  [or $2.18\pm 0.19$]
        \\
  ${\rm arg}(X_A')$
        & $(-138\pm 3)^\circ$ \  [or $(130\pm 3)^\circ$]
\end{tabular}
\end{ruledtabular}
\end{table*}

One can see from Table~\ref{tab:results} that the $D\to PP$ data
can be nicely fitted by the present picture.

\subsection{$D^{+}\rightarrow \pi^{+}\pi^{0}$  vs. $D^+
\to \overline{K}^{0}\pi^{+}$}

Consider the ratio
\begin{eqnarray}
 R_1=2 \bigg|\frac{V_{cs}}{V_{cd}}\bigg|^2
 \frac{\Gamma(D^+\to \pi^+ \pi^0)}{\Gamma(D^+ \to \overline K^0 \pi^+)}.
\end{eqnarray}
The data show $R_1= 3.46\pm 1.17$, whereas $R_1=1$ in
the SU(3) limit.
It is interesting to note that both $D^{+}\rightarrow \pi^{+}\pi^{0}$ and
$D^{+}\rightarrow \overline{K}^{0}\pi^{+}$
amplitudes are identical during SU(3) rescattering
because they do not rescatter with other decay modes.
Moreover, these two amplitudes have no SD weak annihilation
corrections. To take into account the BRs and their ratio
\begin{eqnarray}
R_1= \bigg| \frac{ (a_{1}+a_2)f_{\pi }\left(
m_{D}^{2}-m_{\pi}^{2}\right) F^{D\pi}_0\left(
m_{\pi}^{2}\right)}{a_{1} f_{\pi } (m_{D}^{2}-m_{K}^{2} ) F^{DK}_0
(m_{\pi}^{2})+ a_2 f_{K}\left( m_{D}^{2}-m_{\pi}^{2}\right)
F^{D\pi}_0\left( m_{K}^{2}\right)}\bigg|^2,
\end{eqnarray}
a small $1/N_c^{\rm eff}$ and $F_0^{D\pi}(0)\gsim F_0^{DK}(0)$ are
preferred (see also the discussion in footnote~\ref{footnote:3}).

\subsection{$D^0 \to \pi^+ \pi^-$  vs. $D^0\to K^+ K^-$}

The experiments have measured the ratio
\begin{eqnarray}
 R_2=\frac{\Gamma(D^0\to K^+K^-)}{\Gamma(D^0 \to \pi^+\pi^-)}=2.82\pm
 0.01,
\end{eqnarray}
which is a long-standing puzzle because the conventional
factorization approach yields $R_2=1$ in the SU(3) limit (see
discussions in Ref.~\cite{Cheng:2002wu}). We found that the SD
weak annihilation contributions together with FSIs interfere
destructively to the $D^0 \to \pi^+ \pi^-$ amplitude, but
constructively to the $D^0 \to K^+ K^-$ amplitude, such that the
ratio can be accounted for.

\subsection{$D^0 \to K^0 \overline K^0$}

In the limit of SU(3) symmetry, the $D^0 \to K^0 \overline K^0$
amplitude vanishes. It was explained in Ref.~\cite{Dai:1999cs}
that the non-small branching ratio of this mode may be owing to
long-distance FSIs. Nevertheless, here we conclude that $D\to
\overline K^0 K^0$ occurs mainly due to nonzero SD weak
annihilation effects originating from SU(3) symmetry-breaking
corrections to the distribution amplitudes of the
kaons.\footnote{In spirit, our conclusion agrees with the result
in Ref.~\cite{Eeg:2001un}, where the authors used the chiral
perturbation theory to calculate the weak annihilation effects and
found that the result is proportional to $m_s$.} Moreover we find
$a_1^K=-0.15^{+0.00}_{-0.01}$ in the best fit, which is consistent
with the result given in Ref.~\cite{Ball:2003sc} but in contrast
with that in Ref.~\cite{Braun:2004vf} where the value is
positive.\footnote{\label{footnote:3}There also exists a solution
of positive $a_1^K\simeq 0.19$, $\delta\simeq -39^\circ,
\sigma\simeq -12^\circ, N_c^{\rm eff}\simeq -14,
X_A=2.7e^{i92^\circ}, X_A'=1.7e^{-i104^\circ}$ [or
$X_A'=3.5e^{i152^\circ}$] and $F_0^{D\pi}(0)/F_0^{DK}(0) \simeq
1.05$ with a larger $\chi_{\rm min}\simeq 11.4$.} Note that it has
been argued in Refs.~\cite{Braun:2004vf,Khodjamirian:2004ga} that
the result given in Ref.~\cite{Ball:2003sc} is less reliable.

\subsection{$D$ decays involving $\eta$ or $\eta'$}

It should be stressed that the SD weak annihilation and SU(3)
rescattering effects enter the amplitudes in different ways. For
instance, ${\cal B}(D^0\to \overline K^0\eta)$ and ${\cal B}(D^+
\to \pi^+\eta)$ are mainly enhanced by SU(3) rescattering, whereas
${\cal B}(D^0\to \overline K^0\eta')$ receives contributions
mainly from the SD weak annihilation. This mechanism can be
further tested experimentally from the relative values of the $D^0
\to \pi^0\eta, \pi^0 \eta',\eta\eta$ and $\eta\eta'$ branching
ratios.

\vspace{0.5cm}

\section{Discussions and summary}\label{sec:summary}

\begin{figure}[t]
\vspace{0cm} \centerline{\epsfig{file=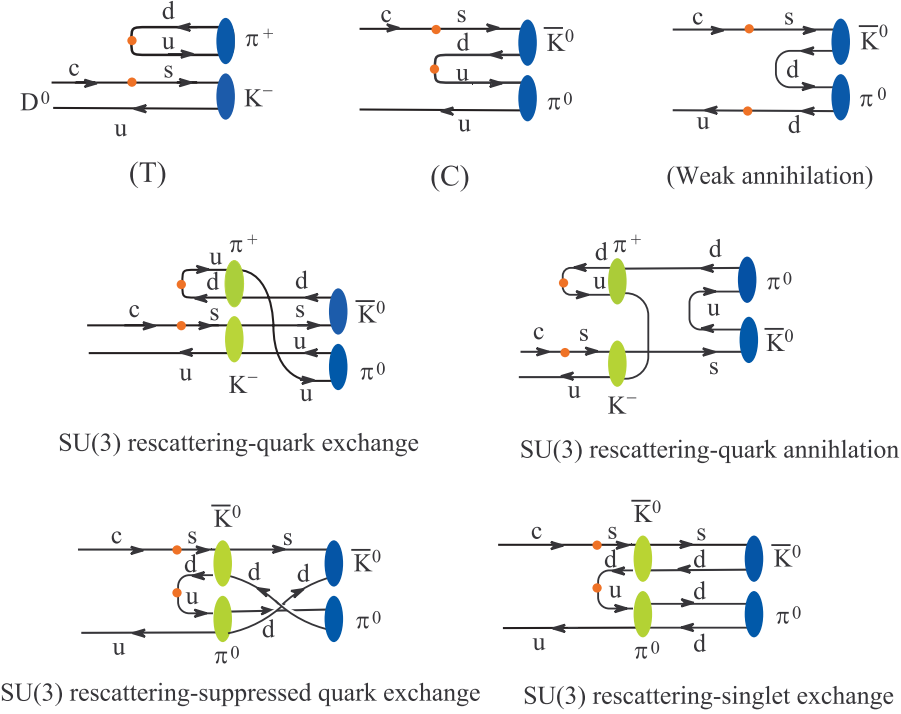,clip=46mm,
width=13cm} } \vspace{0cm} \caption{\small Topologies relevant to
$D\to K \pi$. The second and third rows correspond to the
long-distance SU(3) rescattering contributions to $D^0\to
\overline K^0 \pi^0$ originating from the tree amplitude, where
the quark exchange and singlet exchange contribute to $C$, the
suppressed quark exchange to $T$, and the quark annihilation to
the weak annihilation. The dots denote the quark fields contained
in $(V-A)\otimes (V-A)$ four-quark operators.}
\label{fig:topologies}
\end{figure}

 We have built up a simple model that
the $D\to PP$ decay processes go through ``{\it bare}" amplitudes
followed by elastic-SU(3) rescatterings, where the {\it bare}
amplitude consists of (i) the usual factorization amplitudes of
color-allowance and color-suppression, discussed in
Sec.~\ref{subsec:FacAmp}, and (ii) the SD weak annihilation
amplitudes ($W$-exchange and/or $W$-annihilation) presented in
Sec.~\ref{subsec:ann}. A similar model estimate was proposed by
Chernyak and Zhitnisky~\cite{Chernyak:1981zz}, who considered the
{\it bare amplitude} followed by SU(2) rescattering for $D\to
K\pi$.

In terms of quark-graph amplitudes in the diagrammatic
approach~\cite{Chau:1982da,Chau:1986jb,Chau:1987tk,Rosner:1999xd,Neubert:1997wb,Chiang:2002mr},
the topologies relevant to $D\to PP$ decays are the tree topology
$``T"$, the color-suppressed tree topology $``C"$, and weak
annihilation topologies ($W$-exchange and/or $W$-annihilation),
shown in the first row of Fig.~\ref{fig:topologies}. It has been
stressed in Ref.~\cite{Neubert:1997wb} that in the diagrammatic
approach even though the SD weak annihilation contributions are
neglected, it is still possible for that the weak annihilation
topologies receive sizable contributions from the LD final-state
rescatterings of the (color-suppressed) tree amplitude $T$ ($C$),
as sketched in the second and third rows of
Fig.~\ref{fig:topologies} (some estimates for LD effects see
Refs.~\cite{Gronau:1999zt,Ablikim:2002ep,Cheng:2002wu,Elaaoud:1999pj}).
Moreover, it should be stressed that the SD weak annihilation
amplitudes $A_{\cal T_B}$ have sizable magnitudes comparable to
the factorization amplitudes ${\cal A_{\rm fac}}$; due to the
structure of $(V-A)\otimes(V-A)$ operators in the weak Hamiltonian
relevant to the $D$ decays, as given in (\ref{eq:blocks}) the SD
annihilation contributions are dominated by the topologies of
gluon emission arising from the initial state quarks of the weak
vertex, whereas the contributions vanish in order of $\alpha_s$ if
the gluon is emitted from the final state quarks, i.e., the
amplitudes drawn in Figs.~\ref{fig:ann}(a) and (b) cancel each
other.

We expect that the possible double counting is reasonably
negligible between the LD rescatterings and SD weak annihilation
amplitudes due to the following three reasons: (i) The LD
rescatterings mainly contain {\it gluon exchanges between the two
final-state mesons}, as depicted in Fig.~\ref{fig:topologies},
while the gluon emission originating from the {\it initial-state
quarks} of the weak vertex gives rise to the nonzero SD weak
annihilation amplitudes, as shown in Figs.~\ref{fig:ann}(c) and
(d). (ii) The LD FSIs are dominated by rescatterings of the
(color-suppressed) tree amplitudes which are quite different from
the mechanism of the SD weak annihilation amplitudes. (iii) The LD
rescattering and SD weak annihilation contribute to amplitudes in
different ways; for instance, as seen explicitly in
Table~\ref{tab:results}, the LS rescattering (SD weak
annihilation) interfere constructively (destructively) in the the
$D^+\to K^0\pi^+$ and $D^0\to \pi^0\pi^0$ amplitudes. Finally, it
should be noted that, in $B$ decays, one may worry the double
counting problem since the nonzero weak annihilation is due to the
gluon attached to the {\it final-state quarks} in the
$2(S-P)\otimes(S+P)$ weak vertex.

The strong phase can be generated from the radiative corrections
to the weak vertex and the spectator interactions. Such effects
were lumped into $N_c^{\rm eff}$ as we calculated the
factorization amplitudes. However, since the magnitude of
$1/N_c^{\rm eff}$ is very small obtained in our analysis, it is
thus reasonable to neglect the resulting strong phase; choosing a
real number of $N_c^{\rm eff}$, we have a very nice fit since
$\chi^2_{\rm min}/{\rm d.o.f.}=(5.3^{+1.3}_{-0.5})/5$ for negative
$a_1^K$ (or $\simeq 11.4/5$ for positive $a_1^K$). In other words,
the LD rescattering effects should be approximately absent from
``$N_c^{\rm eff}$''.

Our remaining results are briefly summarized as follows.
\begin{itemize}

\item The two modest rescattering phase differences are
$\delta\equiv \delta_{27}-\delta_{8}\simeq -46^\circ$ and
$\sigma\equiv \delta_{27}-\delta_{1}\simeq -21^\circ$, where the
$\sigma$ phase enters only in the
$\pi^+\pi^-$-$\pi^0\pi^0$-$K^+K^-$-$K^0\overline
K^0$-$\pi^0\eta_8$-$\eta_8\eta_8$ rescattering subset.

\item  We obtain the weak annihilation parameter $|X_A| = 3.84\pm
0.06$ [$|X_A'| = 2.45^{+0.07}_{-0.46}$ or $2.18\pm 0.19$] with a
large phase $(-138\pm 3)^\circ$ [$(-138\pm 3)^\circ$ or $(130\pm
3)^\circ$], where a twofold solution exists for $X_A'$.

\item The $D^0 \to K^0 \overline K^0$ decay occurs mainly due to
the short-distance weak annihilation effects, arising from SU(3)
symmetry-breaking corrections to the distribution amplitudes of
the final-state kaons, but receives negligible contributions from
other modes via SU(3) rescattering.

\item Our results are in good agreement with the experimental
measurements. The predictions for the branching ratios of some
unmeasured modes can be used to test our model in the near future.

\end{itemize}

\acknowledgments We are grateful to Chuan-Hung Chen, Hai-Yang
Cheng, Dao-Neng Gao, and Chien-Wen Hwang  for useful discussions.
This work was supported in part by the National Science Council of
R.O.C. under Grant Nos. NSC93-2112-M-033-004 and
NSC94-2112-M-033-001.

\appendix

\section{Factorization amplitudes}\label{app:fac}

\begin{eqnarray*}
{\cal A}_{\rm fac} ( D^{0}\rightarrow K^{-}\pi^{+})
&=&i\frac{G_{F}}{\sqrt{2}} V_{ud}V_{cs}^{\ast }a_{1}f_{\pi }
(m_{D}^{2}-m_{K}^{2}) F^{DK}_0 (m_{\pi}^{2}), \\
{\cal A}_{\rm fac} (D^{0}\rightarrow \overline{K}^{0}\pi^{0}) &=&
i\frac{G_{F}}{2} V_{ud}V_{cs}^{\ast }a_{2} f_{K} (m_{D}^{2}-m_{\pi
}^{2}) F^{D\pi}_0
(m_{K}^{2}), \\
{\cal A}_{\rm fac} (D^{0}\rightarrow \overline{K}^{0}\eta_{8}) &=&
i \frac{G_{F}}{\sqrt{2}}V_{ud}V_{cs}^{\ast }a_{2} f_{K}
[\cos\vartheta(m_{D}^{2}-m_{\eta}^{2}) F^{D\eta}_0 (m_{K}^{2}) \nonumber\\
&&+ \sin\vartheta(m_{D}^{2}-m_{\eta'}^{2}) F^{D\eta'}_0 (m_{K}^{2})],\\
{\cal A}_{\rm fac} (D^{0}\rightarrow \overline{K}^{0}\eta_{0}) &=&
i\frac{G_{F}}{\sqrt{2}}V_{ud}V_{cs}^{\ast }a_{2} f_{K}
[-\sin\vartheta(m_{D}^{2}-m_{\eta}^{2}) F^{D\eta}_0 (m_{K}^{2}) \nonumber\\
&&+ \cos\vartheta(m_{D}^{2}-m_{\eta'}^{2}) F^{D\eta'}_0 (m_{K}^{2})],\\
\end{eqnarray*}
\begin{eqnarray*}
{\cal A}_{\rm fac}(D^{+}\rightarrow K^{0}\pi^{+}) &=&
i\frac{G_{F}}{\sqrt{2}} V_{us}V_{cd}^{\ast }a_{2} f_{K} (
m_{D}^{2}-m_{\pi}^{2} )
F^{D\pi}_0 ( m_{K}^{2}), \\
{\cal A}_{\rm fac} ( D^{+}\rightarrow K^{+}\pi^{0})
&=&-i\frac{G_{F}}{2} V_{us}V_{cd}^{\ast }a_{1} f_{K}(
m_{D}^{2}-m_{\pi}^{2}) F^{D\pi}_0 ( m_{K}^{2} )\\
{\cal A}_{\rm fac} ( D^{+}\rightarrow K^{+}\eta_{8} ) &=&
i\frac{G_{F}}{\sqrt{2}} V_{us}V_{cd}^{\ast }a_{1} f_{K}
[\cos\vartheta(m_{D}^{2}-m_{\eta}^{2}) F^{D\eta}_0 (m_{K}^{2}) \nonumber\\
&&+ \sin\vartheta(m_{D}^{2}-m_{\eta'}^{2}) F^{D\eta'}_0 (m_{K}^{2})],\\
{\cal A}_{\rm fac} ( D^{+}\rightarrow K^{+}\eta_{0} ) &=&
i\frac{G_{F}}{\sqrt{2}} V_{us}V_{cd}^{\ast }a_{1} f_{K}
[-\sin\vartheta(m_{D}^{2}-m_{\eta}^{2}) F^{D\eta}_0 (m_{K}^{2}) \nonumber\\
&&+ \cos\vartheta(m_{D}^{2}-m_{\eta'}^{2}) F^{D\eta'}_0 (m_{K}^{2})],\\
\end{eqnarray*}
\begin{eqnarray*}
{\cal A}_{\rm fac} ( D^{+}\rightarrow \overline{K}^{0}\pi^{+}) &=&
i \frac{G_{F}}{\sqrt{2}}V_{ud}V_{cs}^{\ast }
[a_{1} f_{\pi}(m_{D}^{2} -m_{K}^{2}) F^{DK}_0( m_{\pi}^{2} ) \\
&&+a_{2}if_{K} ( m_{D}^{2}-m_{\pi}^{2} ) F^{D\pi}_0 (m_K^{2}) ],
\end{eqnarray*}
\begin{eqnarray*}
{\cal A}_{\rm fac} (D^{0}\rightarrow K^{+}\pi^{-}) &=&
i\frac{G_{F}}{\sqrt{2}} V_{us}V_{cd}^{\ast }a_{1} f_{K}
(m_{D}^{2}-m_{\pi}^{2}) F^{D\pi}_0 (m_{K}^{2}), \\
{\cal A}_{\rm fac} (D^{0}\rightarrow K^{0}\pi^0) &=&
i\frac{G_{F}}{2} V_{us}V_{cd}^{\ast }a_{2} f_{K}
(m_{D}^{2}-m_{\pi}^{2})
F^{D\pi}_0 (m_{K^{0}}^{2}), \\
{\cal A}_{\rm fac} (D^{0}\rightarrow K^{0}\eta_{8}) &=&
i\frac{G_{F}}{\sqrt{2}} V_{us}V_{cd}^{\ast }a_{2} f_{K}
[\cos\vartheta(m_{D}^{2}-m_{\eta}^{2}) F^{D\eta}_0 (m_{K}^{2}) \nonumber\\
&&+ \sin\vartheta(m_{D}^{2}-m_{\eta'}^{2}) F^{D\eta'}_0 (m_{K}^{2})],\\
{\cal A}_{\rm fac} (D^{0}\rightarrow K^{0}\eta_{0}) &=&
i\frac{G_{F}}{\sqrt{2}} V_{us}V_{cd}^{\ast }a_{2} f_{K}
[-\sin\vartheta(m_{D}^{2}-m_{\eta}^{2}) F^{D\eta}_0 (m_{K}^{2}) \nonumber\\
&&+ \cos\vartheta(m_{D}^{2}-m_{\eta'}^{2}) F^{D\eta'}_0 (m_{K}^{2})],\\
\end{eqnarray*}
\begin{eqnarray*}
{\cal A}_{\rm fac} (D^{+}\rightarrow \pi^{+}\pi^{0}) &=&-
i\frac{G_{F}}{2} V_{ud}V_{cd}^{*} (a_{1}+a_2) f_{\pi }
 (m_{D}^{2}-m_{\pi}^{2}) F^{D\pi}_0 ( m_{\pi }^{2}), \\
{\cal A}_{\rm fac} (D^{+}\rightarrow \pi^{+}\eta_{8}) &=&
i\frac{G_{F}}{\sqrt{2}} V_{ud}V_{cd}^{\ast}\bigg\{ a_{1} f_{\pi}
[\cos\vartheta(m_{D}^{2}-m_{\eta}^{2}) F^{D\eta}_0 (m_{\pi}^{2}) \nonumber\\
&&+ \sin\vartheta(m_{D}^{2}-m_{\eta'}^{2}) F^{D\eta'}_0
(m_{\pi}^{2})]  \\
&&+  a_2 [f_\eta^u\cos\vartheta(m_{D}^{2}-m_{\pi}^{2}) F^{D\pi}_0
(m_{\eta}^{2}) \\
&& + f_{\eta'}^u\sin\vartheta(m_{D}^{2}-m_{\pi}^{2})
 F^{D\pi}_0 (m_{\eta'}^{2})]\bigg\} \nonumber \\
&&+ i \frac{G_{F}}{\sqrt{2}}  V_{us}V_{cs}^{*} a_2
[f_\eta^s\cos\vartheta(m_{D}^{2}-m_{\pi}^{2}) F^{D\pi}_0
(m_{\eta}^{2})\\
&& + f_{\eta'}^s\sin\vartheta(m_{D}^{2}-m_{\pi}^{2}) F^{D\pi}_0
(m_{\eta'}^{2})], \nonumber \\
 {\cal A}_{\rm fac} (D^{+}\rightarrow
K^{+}\overline{K}^{0}) &=& i
\frac{G_{F}}{\sqrt{2}}V_{us}V_{cs}^{*} a_{1} f_{K}
(m_{D}^{2}-m_K^{2})  F^{DK}_0 (m_{K}^{2}),\\
{\cal A}_{\rm fac} (D^{+}\rightarrow \pi^{+}\eta_{0}) &=&
i\frac{G_{F}}{\sqrt{2}} V_{ud}V_{cd}^{\ast}\bigg\{ a_{1} f_{\pi}
[-\sin\vartheta(m_{D}^{2}-m_{\eta}^{2}) F^{D\eta}_0 (m_{\pi}^{2}) \nonumber\\
&&+ \cos\vartheta(m_{D}^{2}-m_{\eta'}^{2}) F^{D\eta'}_0
(m_{\pi}^{2})]  \\
&&+ a_2 [-f_\eta^u\sin\vartheta(m_{D}^{2}-m_{\pi}^{2}) F^{D\pi}_0
(m_{\eta}^{2})\\
&& + f_{\eta'}^u\cos\vartheta(m_{D}^{2}-m_{\pi}^{2}) F^{D\pi}_0 (m_{\eta'}^{2})\bigg\} \nonumber \\
&&+ i \frac{G_{F}}{\sqrt{2}}  V_{us}V_{cs}^{*} a_2
[-f_\eta^s\sin\vartheta(m_{D}^{2}-m_{\pi}^{2}) F^{D\pi}_0
(m_{\eta}^{2})\\
&& + f_{\eta'}^s\cos\vartheta(m_{D}^{2}-m_{\pi}^{2}) F^{D\pi}_0
(m_{\eta'}^{2})],
\end{eqnarray*}
\begin{eqnarray}
{\cal A}_{\rm fac} (D^{0}\rightarrow \pi^{+}\pi^{-}) &=& i
\frac{G_{F}}{\sqrt{2}} V_{ud}V_{cd}^{\ast }a_{1} f_{\pi }
(m_{D}^{2}-m_{\pi}^{2}) F^{D\pi}_0 (m_{\pi}^{2}),  \nonumber \\
{\cal A}_{\rm fac} ( D^{0}\rightarrow \pi^{0}\pi^{0}) &=&-i
\frac{G_{F}}{2} V_{ud}V_{cd}^{\ast }a_{2} f_{\pi }
(m_{D}^{2}-m_{\pi}^{2}) F^{D\pi }_0 (m_{\pi}^{2}),  \nonumber \\
{\cal A}_{\rm fac} (D^{0}\rightarrow \eta_8 \eta_{8}) &=&
i G_{F} V_{ud}V_{cd}^{*}a_{2} \nonumber\\
&& \hspace{-0.5cm} \times \bigg\{
 f_{\eta}^u[ \cos^2\vartheta(m_{D}^{2}-m_{\eta}^{2}) F^{D\eta}
(m_{\eta}^{2}) + \sin\theta \cos\vartheta(m_{D}^{2}-m_{\eta'}^{2})
F^{D\eta'}_0 (m_{\eta}^{2})] \nonumber\\
&& \hspace{-0.5cm} + f_{\eta'}^u [
\sin^2\vartheta(m_{D}^{2}-m_{\eta'}^{2}) F^{D\eta'}_0
(m_{\eta'}^{2}) +
\sin\vartheta\cos\vartheta(m_{D}^{2}-m_{\eta}^{2})
 F^{D\eta}_0 (m_{\eta'}^{2})]\bigg\}, \nonumber \\
{\cal A}_{\rm fac} (D^{0}\rightarrow K^{+}K^{-}) &=& i
\frac{G_{F}}{\sqrt{2}} V_{us}V_{cs}^* a_{1} f_{K}
(m_{D}^{2}-m_{K}^{2})
F^{DK}_0 (m_{K}^{2}),  \nonumber \\
{\cal A}_{\rm fac} (D^{0}\rightarrow K^{0}\overline{K}^{0})
&=& 0,  \nonumber \\
{\cal A}_{\rm fac} (D^{0}\rightarrow \pi^{0}\eta_{8}) &=&
i\frac{G_{F}}{2} V_{ud}V_{cd}^{*}a_{2} \nonumber\\
&& \times \bigg\{-f_{\pi} [\cos\vartheta(m_{D}^{2}-m_{\eta}^{2})
F^{D\eta}_0 (m_{\pi}^{2}) + \sin\vartheta(m_{D}^{2}-m_{\eta'}^{2})
F^{D\eta'}_0
(m_{\pi}^{2})] \nonumber\\
&&+  [f_\eta^u\cos\vartheta(m_{D}^{2}-m_{\pi}^{2}) F^{D\pi}_0
(m_{\eta}^{2}) + f_{\eta'}^u\sin\vartheta(m_{D}^{2}-m_{\pi}^{2})
 F^{D\pi}_0 (m_{\eta'}^{2})]\bigg\}, \nonumber \\
{\cal A}_{\rm fac} (D^{0}\rightarrow \pi^{0}\eta_{0}) &=&
i\frac{G_{F}}{2} V_{ud}V_{cd}^{*}a_{2} \nonumber\\
&& \hspace{-0.3cm} \times \bigg\{ -f_{\pi}
[-\sin\vartheta(m_{D}^{2}-m_{\eta}^{2}) F^{D\eta}_0 (m_{\pi}^{2})
+ \cos\vartheta(m_{D}^{2}-m_{\eta'}^{2}) F^{D\eta'}_0
(m_{\pi}^{2})] \nonumber\\
&&\hspace{-0.3cm} + [-f_\eta^u\sin\vartheta(m_{D}^{2}-m_{\pi}^{2})
F^{D\pi}_0 (m_{\eta}^{2}) +
f_{\eta'}^u\cos\vartheta(m_{D}^{2}-m_{\pi}^{2})
 F^{D\pi}_0 (m_{\eta'}^{2})]\bigg\}, \nonumber \\
{\cal A}_{\rm fac} (D^{0}\rightarrow \eta_8 \eta_{0}) &=&
i \sqrt{2} G_{F} V_{ud}V_{cd}^{*}a_{2} \nonumber\\
&& \hspace{-0.3cm}\times \bigg\{
 f_{\eta}^u[ -\sin 2\vartheta(m_{D}^{2}-m_{\eta}^{2}) F^{D\eta}
(m_{\eta}^{2}) + \cos 2\theta(m_{D}^{2}-m_{\eta'}^{2})
F^{D\eta'}_0 (m_{\eta}^{2})] \nonumber\\
&&\hspace{-0.3cm} + f_{\eta'}^u [ -\sin
2\vartheta(m_{D}^{2}-m_{\eta'}^{2}) F^{D\eta'}_0 (m_{\eta'}^{2}) +
\cos 2\vartheta(m_{D}^{2}-m_{\eta}^{2})
 F^{D\eta}_0 (m_{\eta'}^{2})]\bigg\}, \nonumber \\
{\cal A}_{\rm fac} (D^{0}\rightarrow \eta_0 \eta_{0}) &=&
i G_{F} V_{ud}V_{cd}^{*}a_{2} \nonumber\\
&& \hspace{-0.5cm}\times \bigg\{
 f_{\eta}^u[ \sin^2\vartheta(m_{D}^{2}-m_{\eta}^{2}) F^{D\eta}
(m_{\eta}^{2}) - \sin\theta \cos\vartheta(m_{D}^{2}-m_{\eta'}^{2})
F^{D\eta'}_0 (m_{\eta}^{2})] \nonumber\\
&&\hspace{-0.5cm} + f_{\eta'}^u [
\cos^2\vartheta(m_{D}^{2}-m_{\eta'}^{2}) F^{D\eta'}_0
(m_{\eta'}^{2}) -
\sin\vartheta\cos\vartheta(m_{D}^{2}-m_{\eta}^{2})
 F^{D\eta}_0 (m_{\eta'}^{2})]\bigg\}. \nonumber \\
 \label{appeq:fac-amp}
\end{eqnarray}

\section{Weak annihilation amplitudes}\label{app:ann}

Here the basic building blocks for annihilation amplitudes
corresponding to Fig.~\ref{fig:ann} are denoted as $A_1^{i(f)}
(P_2\, P_1)$, where the superscript $i \,(f)$ indicates gluon
emission from the initial (final) state quarks, and $P_2$ ($P_1$)
contains a quark (antiquark) arising from the weak vertex with
longitudinal momentum fraction $x$ and $\bar y$, respectively, so
that the building blocks read
\begin{eqnarray}
 A_1^i (P_2\, P_1)&=& \pi\alpha_s \int_0^1\!
dx dy\,
    \Bigg[ \Phi_{P_2}(x)\,\Phi_{P_1}(y)
    \left( \frac{1}{y(1-x\bar y)} + \frac{1}{\bar x^2 y}
    \right)\nonumber\\
 && + r_{\chi}^{P_2}  r_{\chi}^{P_1}\,\Phi_{P_2}^p(x)\,\Phi_{P_1}^p(y)\,
     \frac{2}{\bar x y} \Bigg], \\
 A_1^f (P_2\, P_1)&=& 0.
\end{eqnarray}
$A_1^i$ can be further expressed in terms of $X_A$ as follows:
\begin{eqnarray}\label{appeq:ais}
   A_1^i &=& \left\{
    \begin{array}{ll}
     \pi\alpha_s\Big[18(X_A-4 +\frac{\pi^2}{3})
   + 2 r_{\chi}^K r_{\chi}^{P_1} X_A^2 \\
\hspace{0.5cm}    + 54 a_1^K (X_A + \frac{4}{3}-
\frac{\pi^2}{3})\Big]\,;
      & \mbox{if~} P_2=K^-, \overline K^0, P_1=\pi, \eta_{8,0}\,, \\
     \pi\alpha_s\Big[18(X_A-4 +\frac{\pi^2}{3})
  + 2 r_{\chi}^K r_{\chi}^{P_2} X_A^2   \\
\hspace{0.5cm} + 18 a_1^K (X_A +29 -3\pi^2 )\Big]\,;
      & \mbox{if~} P_2=\pi, \eta_{8,0}, P_1=K^+, K^0\,, \\
      \pi\alpha_s\Big[18(X_A-4 +\frac{\pi^2}{3})
  + 2 r_{\chi}^K r_{\chi}^{P_1} X_A^2 \\
\hspace{0.5cm}   - 54 a_1^K (X_A + \frac{4}{3} -
\frac{\pi^2}{3})\Big] \,;
      & \mbox{if~}  P_2=K^+,  K^0, P_1=\pi, \eta_{8,0} \,, \\
      \pi\alpha_s\Big[18(X_A-4 +\frac{\pi^2}{3})
  + 2 r_{\chi}^{P_2} r_{\chi}^{P_1} X_A^2 \Big] \,;
      & \mbox{if~} P_2, P_1=\pi, \eta_{8,0}  \,, \\
      \pi\alpha_s\Big[18(X_A-4 +\frac{\pi^2}{3})
  + 2 r_{\chi}^K r_{\chi}^{P_2} X_A^2 \\
\hspace{0.5cm} - 18 a_1^K (X_A +29 -3\pi^2 )\Big]\,;
      & \mbox{if~} P_2=\pi, \eta_{8,0}, P_1=K^-, \overline K^0 \,, \\
      \pi\alpha_s\Big[18(X_A-4 +\frac{\pi^2}{3})
  + 2 (r_{\chi}^K)^2 X_A^2\\
\hspace{0.5cm} - 18 a_1^K (4 X_A + 33 -4\pi^2 )\\
\hspace{0.5cm}   +54 (a_1^{K})^{2}(X_A -71 +7 \pi^2 ) \Big]\,;
      & \mbox{if~} P_2=K^{+,0}, P_1=\overline K^0 \,, \\
      \pi\alpha_s\Big[18(X_A-4 +\frac{\pi^2}{3})
  + 2 (r_{\chi}^K)^2 X_A^2\\
\hspace{0.5cm}   + 18 a_1^K (4 X_A + 33 -4\pi^2 )\\
\hspace{0.5cm}   +54 (a_1^{K})^{2}(X_A -71 +7 \pi^2 ) \Big]\,;
      & \mbox{if~} P_2=K^-(\overline K^0), P_1=K^+(K^0) \,,
    \end{array}\right.
\end{eqnarray}
where $X_A \to X_A'$ for processes containing $\eta_0$.
The complete weak annihilation amplitudes are given by
\begin{eqnarray*}
A_{\cal T_B} (K^- \pi^+) &=&
  i \frac{G_F}{\sqrt{2} } f_D f_{\pi} f_K
  \frac{C_F}{N_c^2} c_{1}  V_{ud} V_{cs}^*
  A^i_1(K^-\, \pi^+),\\
A_{\cal T_B} (\overline K^0 \pi^0)
  &=& -i \frac{G_F}{2 } f_D f_{\pi} f_K
  \frac{C_F}{N_c^2} c_{1}  V_{ud} V_{cs}^*
  A^i_1(\overline K^0\, \pi^0),\\
A_{\cal T_B} (\overline K^0 \eta_8) &=&
  i \frac{G_F}{2 \sqrt 3} f_D f_{\pi} f_{\eta_8}
  \frac{C_F}{N_c^2} c_{1}  V_{ud} V_{cs}^*
  \{-2A^i_1(\eta_8\, \overline K^0) + A^i_1(\overline K^0\, \eta_8)
  \},\\
A_{\cal T_B} (\overline K^0 \eta_0) &=&
  i \frac{G_F}{\sqrt 6} f_D f_{\pi} f_{\eta_0}
  \frac{C_F}{N_c^2} c_{1}  V_{ud} V_{cs}^*
  \{A^i_1(\eta_0\, \overline K^0) + A^i_1(\overline K^0\, \eta_0) \},
\end{eqnarray*}
\begin{eqnarray*}
A_{\cal T_B} (K^0 \pi^+) &=&
  i \frac{G_F}{\sqrt{2} } f_D f_{\pi} f_K
  \frac{C_F}{N_c^2} c_2  V_{us} V_{cd}^*
  A^i_1(\pi^+\, K^0),\\
A_{\cal T_B} (K^+ \pi^0)
  &=&  i \frac{G_F}{2 } f_D f_{\pi} f_K
  \frac{C_F}{N_c^2} c_{2}  V_{us} V_{cd}^*
  A^i_1(\pi^0\, K^+),\\
A_{\cal T_B} (K^+ \eta_8) &=&
  i \frac{G_F}{2 \sqrt 3} f_D f_{\pi} f_{\eta_8}
  \frac{C_F}{N_c^2} c_{2}  V_{us} V_{cd}^*
  \{A^i_1(\eta_8\, K^+)-2 A^i_1(K^+\, \eta_8) \},\\
A_{\cal T_B} (K^+ \eta_0) &=&
  i \frac{G_F}{\sqrt 6} f_D f_{\pi} f_{\eta_0}
  \frac{C_F}{N_c^2} c_{2}  V_{us} V_{cd}^*
  \{A^i_1(\eta_0\, K^+) + A^i_1(K^+\, \eta_0) \},
\end{eqnarray*}
\begin{eqnarray*}
A_{\cal T_B} (\overline K^0 \pi^+)=0, {\hspace{8.2cm}}\\
\end{eqnarray*}
\begin{eqnarray*}
A_{\cal T_B} (K^+ \pi^-)
  &=& i \frac{G_F}{\sqrt{2} } f_D f_{\pi} f_K
  \frac{C_F}{N_c^2} c_{1}  V_{us} V_{cd}^*
  A^i_1(\pi^-\, K^+),\\
A_{\cal T_B} (K^0 \pi^0)
  &=& -i \frac{G_F}{2 } f_D f_{\pi} f_K
  \frac{C_F}{N_c^2} c_{1}  V_{us} V_{cd}^*
  A^i_1(\pi^0\, K^0),\\
A_{\cal T_B} (K^0 \eta_8)
  &=& i \frac{G_F}{2 \sqrt 3} f_D f_{\pi} f_{\eta_8}
  \frac{C_F}{N_c^2} c_{1}  V_{us} V_{cd}^*
  \{A^i_1(\eta_8\, K^0)-2 A^i_1(K^0\, \eta_8) \},\\
A_{\cal T_B} (K^0 \eta_0) &=&
  i \frac{G_F}{\sqrt 6} f_D f_{\pi} f_{\eta_0}
  \frac{C_F}{N_c^2} c_{1}  V_{us} V_{cd}^*
  \{A^i_1(\eta_0\, K^0) + A^i_1(K^0\, \eta_0) \},
\end{eqnarray*}
\begin{eqnarray*}
A_{\cal T_B} (\pi^+ \pi^0)
  &=& i \frac{G_F}{2} f_D f_{\pi}^2
  \frac{C_F}{N_c^2} c_{2}  V_{ud} V_{cd}^*
  \{A^i_1(\pi^0\, \pi^+) - A^i_1(\pi^+\, \pi^0)\} = 0 ,\\
A_{\cal T_B} (\pi^+ \eta_8)
  &=& i \frac{G_F}{2\sqrt 3} f_D f_{\pi} f_{\eta_8}
  \frac{C_F}{N_c^2} c_{2}  V_{ud} V_{cd}^*
  \{A^i_1(\pi^+\, \eta_8) + A^i_1(\eta_8\, \pi^+)\},\\
A_{\cal T_B} (K^+ \overline K^0) &=&
 i \frac{G_F}{\sqrt2} f_D f_{K}^2
  \frac{C_F}{N_c^2} c_{2}  V_{ud} V_{cd}^*  A^i_1(K^+\, \overline
  K^0),\\
A_{\cal T_B} (\pi^+ \eta_0)
  &=& i \frac{G_F}{\sqrt 6} f_D f_{\pi} f_{\eta_0}
  \frac{C_F}{N_c^2} c_{2}  V_{ud} V_{cd}^*
  \{A^i_1(\pi^+\, \eta_0) + A^i_1(\eta_0\, \pi^+)\},\\
\end{eqnarray*}
\begin{eqnarray*}
A_{\cal T_B} (\pi^+ \pi^-)
  &=&  i \frac{G_F}{\sqrt2} f_D f_{\pi}^2
  \frac{C_F}{N_c^2} c_{1}  V_{ud} V_{cd}^*  A^i_1(\pi^-\, \pi^+),\\
A_{\cal T_B} (\pi^0 \pi^0)
  &=& i \frac{G_F}{2} f_D f_{\pi}^2
  \frac{C_F}{N_c^2} c_{1}  V_{ud} V_{cd}^*  A^i_1(\pi^0\, \pi^0),\\
A_{\cal T_B} (\eta_8 \eta_8)
  &=&  i \frac{G_F}{6} f_D f_{\eta_8}^2
  \frac{C_F}{N_c^2} c_{1}
( V_{ud} V_{cd}^* + 4 V_{us} V_{cs}^*) A^i_1(\eta_8\, \eta_8),\\
A_{\cal T_B} (K^+ K^-) &=&
  i \frac{G_F}{\sqrt2} f_D f_{K}^2
  \frac{C_F}{N_c^2} c_{1}  V_{us} V_{cs}^*  A^i_1(K^-\, K^+),\\
A_{\cal T_B} (K^0 \overline K^0) &=&
  i \frac{G_F}{\sqrt2} f_D f_{K}^2
  \frac{C_F}{N_c^2} c_{1}
  \{ V_{us} V_{cs}^* A^i_1(\overline K^0\,K^0)
 +  V_{ud} V_{cd}^* A^i_1(K^0\, \overline K^0)\},\\
A_{\cal T_B} (\pi^0 \eta_8)
  &=& -i \frac{G_F}{2\sqrt6} f_D f_{\pi} f_{\eta_8}
  \frac{C_F}{N_c^2} c_{1}  V_{ud} V_{cd}^*
  \{A^i_1(\pi^0\, \eta_8) + A^i_1(\eta_8\, \pi^0)\}\\
A_{\cal T_B} (\pi^0 \eta_0)
  &=& -i \frac{G_F}{2\sqrt 3} f_D f_{\pi} f_{\eta_0}
  \frac{C_F}{N_c^2} c_{1}  V_{ud} V_{cd}^*
  \{A^i_1(\pi^0\, \eta_0) + A^i_1(\eta_0\, \pi^0)\},\\
A_{\cal T_B} (\eta_8 \eta_0)
  &=& i \frac{G_F}{6} f_D f_{\eta_8} f_{\eta_0}
  \frac{C_F}{N_c^2} c_{1}
  ( V_{ud} V_{cd}^* -2 V_{us} V_{cs}^* )
  \{A^i_1(\eta_0\, \eta_8) + A^i_1(\eta_8\, \eta_0)\},\\
A_{\cal T_B} (\eta_0 \eta_0)
  &=& i \frac{G_F}{3} f_D f_0^2
  \frac{C_F}{N_c^2} c_{1}
  ( V_{ud} V_{cd}^* + V_{us} V_{cs}^* ) A^i_1(\eta_0\, \eta_0) .
 \end{eqnarray*}

The above weak annihilation amplitudes can be further expressed in
terms of $X_A^{(\prime)}$ as follows (in units of $i \frac{G_F}{\sqrt{2} }
f_D f_{P_1} f_{P_2} \frac{C_F}{N_c^2} \pi\alpha_s $):
\begin{eqnarray*}
 A_{\cal T_B} (K^- \pi^+) &=&
  c_{1}  V_{ud} V_{cs}^* \bigg[18(X_A-4 +\frac{\pi^2}{3})
  + 2 r_{\chi}^K r_{\chi}^\pi X_A^2 + 54 a_1^K (X_A +\frac{4}{3}- \frac{\pi^2}{3})\bigg], \\
 A_{\cal T_B} (\overline K^0 \pi^0)
  &=& \frac{1}{\sqrt{2}}A_{\cal T_B} (K^- \pi^+), \\
 A_{\cal T_B} (\overline K^0 \eta_8) &=&
  - \frac{1}{\sqrt{6}}c_{1}  V_{ud} V_{cs}^* \bigg[18(X_A-4 +\frac{\pi^2}{3})
  + 2 r_{\chi}^K r_{\chi}^{\eta_8} X_A^2  - 18 a_1^K (5 X_A + 62 -7\pi^2)
  \bigg],\\
 A_{\cal T_B} (\overline K^0 \eta_0) &=&
   \frac{2}{\sqrt{3}}c_{1}  V_{ud} V_{cs}^* \bigg[18(X_A'-4 +\frac{\pi^2}{3})
  + 2 r_{\chi}^K r_{\chi}^{\eta_0} X_A^{\prime 2} + 9 a_1^K (2 X_A' - 25 +2 \pi^2)\bigg],
\end{eqnarray*}
\begin{eqnarray*}
A_{\cal T_B} (K^0 \pi^+) &=&
  c_{2}  V_{ud} V_{cs}^* \bigg[18(X_A-4 +\frac{\pi^2}{3})
  + 2 r_{\chi}^K r_{\chi}^\pi X_A^2 + 18 a_1^K (X_A +29 -3\pi^2 )\bigg],\\
A_{\cal T_B} (K^+ \pi^0)
  &=& \frac{1}{\sqrt{2}}A_{\cal T_B} (K^0 \pi^+),  \\
A_{\cal T_B} (K^+ \eta_8) &=&
  - \frac{1}{\sqrt{6}}c_{2}  V_{ud} V_{cs}^* \bigg[18(X_A-4 +\frac{\pi^2}{3})
  + 2 r_{\chi}^K r_{\chi}^{\eta_8} X_A^2 - 18 a_1^K (7 X_A + 37
  -5\pi^2)\bigg],\\
A_{\cal T_B} (K^+ \eta_0) &=&
   \frac{2}{\sqrt{3}}c_{2}  V_{ud} V_{cs}^* \bigg[18(X_A'-4 +\frac{\pi^2}{3})
  + 2 r_{\chi}^K r_{\chi}^{\eta_0} X_A^{\prime 2} - 9 a_1^K (2 X_A' - 25 +2 \pi^2)\bigg],
\end{eqnarray*}
\begin{eqnarray*}
A_{\cal T_B} (\overline K^0 \pi^+)=0, {\hspace{7.5cm}}
\end{eqnarray*}
\begin{eqnarray*}
A_{\cal T_B} (K^+ \pi^-)
  &=& \frac{V_{us} V_{cd}^* c_1}{V_{ud} V_{cs}^* c_2} A_{\cal T_B} (K^0 \pi^+), {\hspace{5.cm}}\\
A_{\cal T_B} (K^0 \pi^0)
  &=& -\frac{1}{\sqrt{2}}\frac{V_{us} V_{cd}^* c_1}{V_{ud} V_{cs}^* c_2}A_{\cal T_B} (K^0 \pi^+), \\
A_{\cal T_B} (K^0 \eta_8)
  &=& \frac{V_{us} V_{cd}^* c_1}{V_{ud} V_{cs}^* c_2}A_{\cal T_B} (K^+ \eta_8), \\
A_{\cal T_B} (K^0 \eta_0)
  &=& \frac{V_{us} V_{cd}^* c_1}{V_{ud} V_{cs}^* c_2}A_{\cal T_B} (K^+ \eta_0),
\end{eqnarray*}
\begin{eqnarray*}
A_{\cal T_B} (\pi^+ \pi^0)
  &=& 0 ,\\
A_{\cal T_B} (\pi^+ \eta_8)
  &=&  \sqrt{\frac{2}{3}}c_{2}  V_{ud} V_{cd}^* \bigg[18(X_A-4 +\frac{\pi^2}{3})
  + 2 r_{\chi}^\pi r_{\chi}^{\eta_8} X_A^2 \bigg], \\
A_{\cal T_B} (K^+ \overline K^0) &=&
  c_{2}  V_{ud} V_{cd}^* \bigg[18(X_A-4 +\frac{\pi^2}{3})
  + 2 (r_{\chi}^K)^2 X_A^2 \nonumber\\
  && - 18 a_1^K (4 X_A + 33 -4\pi^2 ) +54 (a_1^K)^2(X_A -71 +7 \pi^2 )
  \bigg],\\
A_{\cal T_B} (\pi^+ \eta_0)
  &=&  \frac{2}{\sqrt{3}}c_{2}  V_{ud} V_{cd}^* \bigg[18(X_A'-4 +\frac{\pi^2}{3})
  + 2 r_{\chi}^\pi r_{\chi}^{\eta_0} X_A^{\prime 2} \bigg],
 \end{eqnarray*}
\begin{eqnarray}
A_{\cal T_B} (\pi^+ \pi^-)
  &=&  c_{1}  V_{ud} V_{cd}^* \bigg[18(X_A-4 +\frac{\pi^2}{3})
  + 2 (r_{\chi}^\pi)^2  X_A^2 \bigg], \nonumber\\
A_{\cal T_B} (\pi^0 \pi^0)
  &=& \frac{1}{\sqrt{2}}A_{\cal T_B} (\pi^+\pi^-  ), \nonumber\\
A_{\cal T_B} (\eta_8 \eta_8)
  &=& - \frac{1}{\sqrt{2}}A_{\cal T_B} (\pi^+\pi^-  ), \nonumber\\
A_{\cal T_B} (K^+ K^-) &=&
  c_{1}  V_{us} V_{cs}^* \bigg[18(X_A-4 +\frac{\pi^2}{3})
  + 2 (r_{\chi}^K)^2 X_A^2 \nonumber\\
  && + 18 a_1^K (4 X_A + 33 -4\pi^2 ) +54 (a_1^K)^2(X_A -71 +7 \pi^2
  )\bigg], \nonumber\\
A_{\cal T_B} (\pi^0 \eta_8)
  &=& -  \frac{1}{\sqrt{3}}c_{1}  V_{ud} V_{cd}^* \bigg[18(X_A-4 +\frac{\pi^2}{3})
  + 2 (r_{\chi}^\pi)^2  X_A^2 \bigg], \nonumber\\
A_{\cal T_B} (\pi^0 \eta_0)
  &=&  - \sqrt{\frac{2}{3}}c_{1}  V_{ud} V_{cd}^* \bigg[18(X_A'-4 +\frac{\pi^2}{3})
  + 2 r_{\chi}^\pi r_{\chi}^{\eta_0} X_A^{\prime 2} \bigg], \nonumber\\
A_{\cal T_B} (\eta_8 \eta_0)
  &=&  - \frac{\sqrt{2}}{3}c_{1}  V_{ud} V_{cd}^* \bigg[18(X_A'-4 +\frac{\pi^2}{3})
  + 2 r_{\chi}^{\eta_8} r_{\chi}^{\eta_0} X_A^{\prime 2} \bigg], \nonumber\\
A_{\cal T_B} (\eta_0 \eta_0)
  &=&  0,
\end{eqnarray}
where $X_A$ is treated as a universal parameter for SU(3)
channels, while for decay modes involving $\eta_0$, it is
distinguished to be $X_A'$.

\section{SU(3) final state interactions --- $ \mathbf{8\otimes 8}$
Decomposition}\label{appsec:fsi}

To describe elastic SU(3) final state interactions among $D\to P_1
P_2$ decays, we adopt the notations:
\begin{equation}
\mathbf{q}=q^{i}=\left(
\begin{array}{c}
q^{1} \\
q^{2} \\
q^{3}
\end{array}
\right) \equiv \left(
\begin{array}{c}
u \\
d \\
s
\end{array}
\right)
\end{equation}
and
\begin{equation}
\overline{\mathbf{q}}=q_{j}=(
\begin{array}{ccc}
q_{1} & q_{2} & q_{3}
\end{array}) \equiv (
\begin{array}{ccc}
\overline{u} & \overline{d} & \overline{s}
\end{array} ) .
\end{equation}
The octet final-state pseudoscalar mesons $P_{1}$\ and $P_{2}$,
which are viewed as composites of quarks in the quark model,  can
be represented by the matrix
\begin{eqnarray}
{\Pi}=\mathbf{q}\otimes\mathbf{\bar q}-\frac{1}{3} \mathbf{1}\
{\rm Tr}(\mathbf{q}\otimes\mathbf{\bar q})= \left(
\begin{array}{ccc}
\frac{\pi^0}{\sqrt2}+\frac{\eta_8}{\sqrt6}&{\pi^+} &{K^+}\\
{\pi^-}&-\frac{\pi^0}{\sqrt2}+ \frac{\eta_8}{\sqrt6}&{K^0}\\
{K^-}  &{\overline K{}^0}&-\sqrt{\frac{2}{3}}{\eta_8}
\end{array}
\right), \label{app:octet}
\end{eqnarray}
where $\Pi_j^i$ is the $\mathbf{8}$ representation, while
$\Pi_i^i=0$. The SU(3) final-state rescatterings for $D\to P_1
P_2$ are described by the product $\mathbf{8\otimes 8}$. Since the
$P_1 P_2$ states obey the Bose symmetry, only the symmetric states
given by the representation $\mathbf{36(=27\oplus 8\oplus 1)}$ in
$\mathbf {8 \otimes 8(= 36 \oplus 28)}$ decomposition are
relevant, whereas states given by the representation
$\mathbf{28(=10\oplus \overline {10}\oplus 8)}$ vanish. The weak
decay amplitudes $\mathbf{A}_i^{\rm FSI}$ for  $D\to P_1 P_2$ with
FSIs are given by
\begin{eqnarray}
{\mathbf{A}_{i}^{\rm FSI}}&=& \sum_l {\mathbf S}^{1/2}_{il}
{\mathbf A_{l}^{\rm bare} }= (\mathbf{U}^{\rm T}{\mathbf
S^{1/2}}_{\rm diag}\,\mathbf{U})_{il} {\mathbf A_{l}^{\rm bare}},
\label{app:fsi}
\end{eqnarray}
where $\mathbf{A}_{l}^{\rm bare} =\mathbf{A}_{l}^{\rm
fac}+\mathbf{A}_{l}^{\cal T_B}$ are defined in
Eqs.~(\ref{eq:subset1.2}), (\ref{eq:subset2.2}),
(\ref{eq:subset3.2}), (\ref{eq:subset4.2}) and
(\ref{eq:subset5.2}). In orthonormal bases of SU(3), the
$\mathbf{S}_{\rm diag}^{1/2}$ matrix, describing the SU(3) FSIs,
can be recast into the following form
\begin{equation}
\mathbf{S}_{\rm diag}^{1/2}=e^{i\delta
_{27}}\underset{a=1}{\overset{27}{\sum }}\left| T\left( 27\right)
;a\right\rangle \left\langle T\left( 27\right) ;a\right|
+e^{i\delta_{8}}\underset{b=1}{\overset{8}{\sum }}\left| T\left(
8\right) ;b\right\rangle \left\langle T\left( 8\right) ;b\right|
+e^{i\delta_{1}}\left| T\left( 1\right) \right\rangle \left\langle
T\left( 1\right) \right| ,
\end{equation}
where $|T(27);a\rangle$, $|T(8);b\rangle$, and $|T(1)\rangle$ are
orthonormal SU(3) bases in the irreducible representation
$\mathbf{36}$. Using the tensor approach~\cite{Lee,georgi}, the
$\mathbf{36}$ states are described by $\Pi^{\{i}_{\ \{k}
\Pi^{j\}}_{\ l\}}$ with $\{i,j\}$ being symmetric in indices
$i,j$, and can be decoupled into three types of irreducible
tensors: (i) $\mathbf{1}$, an irreducible tensor of rank $(0,0)$,
equals to $\Pi^{i}_{\ k} \Pi^{k}_{\ i} \equiv T^{ik}_{ki}$. (ii)
$\mathbf {8}$, an irreducible tensor of rank $(1,1)$, is
equivalent to $T^{mj}_{km}-(1/3)\delta^j_k T^{ml}_{lm}=\Pi^{m}_{\
k} \Pi^{j}_{\ m}-(1/3)\delta^j_k \Pi^{m}_{\ i} \Pi^{i}_{\ m}
\equiv U^{j}_{k}$. (iii) $\mathbf{27}$, an irreducible tensor of
rank $(2,2)$, is given by $T^{ij}_{kl} +
T^{ji}_{kl}-(1/5)(\delta^i_k T^{mj}_{lm}+\delta^j_k
T^{mi}_{lm}+\delta^i_l T^{mj}_{km}+\delta^j_l
T^{mi}_{km})+(1/20)(\delta^i_k \delta^j_l + \delta^j_k
\delta^i_l)T^{mn}_{nm} \equiv V^{ij}_{kl}$. We summarized the
orthonormal states in the representation ${\bf 36}$ together with
their quantum numbers $S$ and $I$ as follows. (i) In the
representation $\bf 1$, the normalized state $\left| T\left(
1\right) \right\rangle $ is
\begin{eqnarray}
(S=0,\,I=0):&& \frac{1}{\sqrt{8}}(\sqrt{2}|\pi^+\pi^-\rangle +
|\pi^0 \pi^0\rangle + |\eta_8 \eta_8\rangle+\sqrt{2}|K^+
K^-\rangle +\sqrt{2}|\overline K^0 K^0\rangle )\,.
\end{eqnarray}
(ii) In the representation $\bf 8$, the normalized states $\left|
T\left( 8\right) ;b\right\rangle$ are
\begin{eqnarray}
(S=1,\,I=\frac{1}{2}):&& \sqrt{\frac{1}{10}}({\sqrt{6}}(|K^0
\pi^+\rangle + \sqrt{3}|K^+ \pi^0\rangle-|K^+\eta_8\rangle),
 \nonumber\\
 && \sqrt{\frac{1}{10}}({\sqrt{6}}(|K^+ \pi^-\rangle
-\sqrt{3}|K^0 \pi^0\rangle -|K^0\eta_8\rangle);
\\
(S=-1,\,I={\frac{1}{2}}):&& \sqrt{\frac{1}{10}}({\sqrt{6}}(|K^-
\pi^+\rangle -\sqrt{3}|\overline K^0 \pi^0\rangle -|\overline
K^0\eta_8\rangle),
 \nonumber\\
 && \sqrt{\frac{1}{10}}(-{\sqrt{6}}(|\overline K^0
\pi^-\rangle - \sqrt{3}|K^- \pi^0\rangle +|K^-\eta_8\rangle);
\\
(S=0,\,I=1):&& \sqrt{\frac{2}{5}}|\pi^+ \eta_8\rangle+
 \sqrt{\frac{3}{5}}|\overline K^0 K^+\rangle ,
 \nonumber\\
 && \sqrt{\frac{1}{10}}(\sqrt{3}|K^+ K^-\rangle
  -\sqrt{3}|\overline K^0 K^0\rangle + 2|\pi^0 \eta_8\rangle), \nonumber\\
&& \sqrt{\frac{2}{5}}|\pi^- \eta_8\rangle
   + \sqrt{\frac{3}{5}}|K^0 K^-\rangle ;\\
(S=0,\,I=0):
 && \sqrt{\frac{1}{10}}(-2|\pi^+\pi^-\rangle
 -\sqrt{2}|\pi^0 \pi^0\rangle + \sqrt{2}|\eta_8 \eta_8\rangle
 +|K^+ K^-\rangle +|\overline K^0 K^0\rangle ).
\end{eqnarray}

(iii) In the representation $\bf 27$, the normalized states
$\left| T\left( 27\right);a\right\rangle$ are
\begin{eqnarray}
(S=2,\,I=1):&& \frac{1}{\sqrt{2}}|K^+K^+\rangle, \quad
|K^+K^0\rangle, \quad \frac{1}{\sqrt{2}}|K^0 K^0\rangle;
\\
(S=-2,\,I=1):&& \frac{1}{\sqrt{2}}|\overline K^0 \overline
K^0\rangle, \quad |\overline K^0 K^-\rangle, \quad
\frac{1}{\sqrt{2}}|K^- K^-\rangle;
\\
(S=1,\,I=\frac{3}{2}):
 &&|K^+ \pi^+\rangle,\quad
  \frac{1}{\sqrt{3}}(|K^0 \pi^+\rangle -\sqrt{2}|K^+\pi^0\rangle),
 \nonumber\\
 && \frac{1}{\sqrt{3}}(|K^+ \pi^-\rangle +\sqrt{2}|K^0\pi^0\rangle),
  \quad |K^0 \pi^-\rangle; \\
(S=-1,\,I=\frac{3}{2}):
 &&|\overline K^0 \pi^+\rangle,\quad
 \frac{1}{\sqrt{3}}(|K^- \pi^+\rangle +\sqrt{2}|\overline K^0 \pi^0\rangle),
 \nonumber\\
 && \frac{1}{\sqrt{3}}(|\overline K^0
\pi^-\rangle -\sqrt{2}|K^-\pi^0\rangle), \quad |K^- \pi^-\rangle;
\\
(S=1,\,I=\frac{1}{2}):&&
\frac{1}{\sqrt{30}}(\sqrt{2}|K^0\pi^+\rangle + |K^+\pi^0\rangle
+3\sqrt3|K^+\eta_8\rangle), \nonumber
\\
&&\frac{1}{\sqrt{30}}(\sqrt{2}|K^+\pi^-\rangle - |K^0\pi^0\rangle
+3\sqrt3|K^0\eta_8\rangle);
\\
(S=-1,\,I=\frac{1}{2}):&&
\frac{1}{\sqrt{30}}(\sqrt{2}|K^-\pi^+\rangle - |\overline
K^0\pi^0\rangle +3\sqrt3|\overline K^0\eta_8\rangle), \nonumber
\\
&&\frac{1}{\sqrt{30}}(\sqrt{2}|\overline K^0\pi^-\rangle +
|K^-\pi^0\rangle +3\sqrt3|K^-\eta_8\rangle);
\\
(S=0,\,I=2):
 && \frac{1}{\sqrt{2}}|\pi^+ \pi^+\rangle, \quad
 |\pi^0 \pi^+\rangle, \quad
 \frac{1}{\sqrt3}(|\pi^- \pi^+\rangle\nonumber\\
 &&
 -\sqrt{2}|\pi^0 \pi^0\rangle, \quad |\pi^0\pi^-\rangle, \quad
  \frac{1}{\sqrt{2}}|\pi^- \pi^-\rangle; \\
(S=0,\,I=1):
 && \sqrt{\frac{2}{5}}|\overline K^0 K^+\rangle
 -\sqrt{\frac{3}{5}}|\pi^+ \eta_8\rangle, \quad
 \nonumber\\
 &&\sqrt{\frac{1}{5}}(|K^+ K^-\rangle -|\overline K^0 K^0\rangle -
 \sqrt{3}|\pi^0 \eta_8\rangle), \nonumber \\
&& \sqrt{\frac{2}{5}}|K^0 K^-\rangle - \sqrt{\frac{3}{5}}|\pi^-
\eta_8\rangle;
\\
(S=0,\,I=0):&& \frac{1}{4\sqrt{15}}(2|\pi^+\pi^-\rangle +
\sqrt{2}|\pi^0 \pi^0\rangle + 9\sqrt{2} |\eta_8 \eta_8\rangle
 - 6|K^+ K^-\rangle -6|\overline K^0 K^0\rangle ). {\hspace{1cm}}
\end{eqnarray}
Using the above results, one can immediately obtain the relevant
$\mathbf{U}$ matrices and the corresponding SU(3) eigen-amplitudes
in $D$ decays:
\begin{eqnarray}
\mathbf{A}_{(\overline{K\pi})^0}^{\rm SU(3)} &=& \left(
\begin{array}{c}
 |\mathbf{27},S=-1,I=3/2,I_z=+1/2 \rangle\\
 |\mathbf{27},S=-1,I=1/2,I_z=+1/2 \rangle \\
 |\mathbf{8},S=-1,I=1/2,I_z=+1/2 \rangle  \end{array} \right)\nonumber\\
 &=& \mathbf{U}_{(\overline{K\pi})^0} \mathbf{A}_{(\overline{K\pi})^0}^{\rm bare} = \left(
\begin{array}{ccc}
\sqrt{\frac{1}{3}} & \sqrt{\frac{2}{3}}
  & 0
  \\
\sqrt{\frac{1}{15}} & -\sqrt{\frac{1}{30}}
  & \frac{3}{\sqrt{10}} \\
\sqrt{\frac{3}{5}}& -\sqrt{\frac{3}{10}}
  & -\sqrt{\frac{1}{10}}
\end{array} \right)\mathbf{A}_{(\overline{K\pi})^0}^{\rm bare},\label{app:su3kpibar0}
\end{eqnarray}
\begin{eqnarray}
\mathbf{A}_{({K\pi})^0}^{\rm SU(3)} &=& \left(
\begin{array}{c}
 |\mathbf{27},S=1,I=3/2,I_z=-1/2 \rangle\\
 |\mathbf{27},S=1,I=1/2,I_z=-1/2 \rangle \\
 |\mathbf{8},S=1,I=1/2,I_z=-1/2 \rangle  \end{array} \right)\nonumber\\
 &=& \mathbf{U}_{({K\pi})^0} \mathbf{A}_{({K\pi})^0}^{\rm bare} = \left(
\begin{array}{ccc}
\sqrt{\frac{1}{3}} & \sqrt{\frac{2}{3}}
  & 0
  \\
\sqrt{\frac{1}{15}} & -\sqrt{\frac{1}{30}}
  & \frac{3}{\sqrt{10}} \\
\sqrt{\frac{3}{5}}& -\sqrt{\frac{3}{10}}
  & -\sqrt{\frac{1}{10}}
\end{array} \right)\mathbf{A}_{({K\pi})^0}^{\rm bare},\label{app:su3kpi0}
\end{eqnarray}
\begin{eqnarray}
\mathbf{A}_{(K\pi)^+}^{\rm SU(3)} &=& \left( \begin{array}{c}
 |\mathbf{27},S=1,I=3/2,I_z=1/2 \rangle  \\
 |\mathbf{27},S=1,I=1/2,I_z=1/2 \rangle \\
 |\mathbf{8},S=1,I=1/2,I_z=1/2 \rangle  \end{array} \right)
 \nonumber\\
 &=&
 \mathbf{U}_{(K\pi)^+} \mathbf{A}_{(K\pi)^+}^{\rm bare}= \left(
\begin{array}{ccc}
\sqrt{\frac{1}{3}} & -\sqrt{\frac{2}{3}}
  & 0
  \\
\sqrt{\frac{1}{15}} & \sqrt{\frac{1}{30}}
  & \frac{3}{\sqrt{10}} \\
  \sqrt{\frac{3}{5}}& \sqrt{\frac{3}{10}}
  & -\sqrt{\frac{1}{10}}
\end{array} \right)\mathbf{A}_{(K\pi)^+}^{\rm bare},\label{app:su3kpip}
\end{eqnarray}
\begin{eqnarray}
\mathbf{A}_{(\pi\pi)^+}^{\rm SU(3)} &=& \left( \begin{array}{c}
 |\mathbf{27},S=0,I=2,I_z=1 \rangle \\
 |\mathbf{27},S=0,I=1,I_z=1 \rangle \\
 |\mathbf{8},S=0,I=1,I_z=1 \rangle \end{array} \right)\nonumber\\
 &=&
 \mathbf{U}_{(\pi\pi)^+} \mathbf{A}_{(\pi\pi)^+}^{\rm bare}= \left(
\begin{array}{ccc}
1&0 & 0
  \\
0 & -\sqrt{\frac{3}{5}}
  & \sqrt{\frac{2}{5}} \\
0& \sqrt{\frac{2}{5}}
  & \sqrt{\frac{3}{5}}
\end{array} \right)\mathbf{A}_{(\pi\pi)^+}^{\rm bare},\label{app:su3pipip}
\end{eqnarray}
\begin{eqnarray}
\mathbf{A}_{(\pi\pi)^0}^{\rm SU(3)}&=&\left( \begin{array}{c}
 |\mathbf{27},S=0,I=2,I_z=0 \rangle\\
 |\mathbf{27},S=0,I=0,I_z=0 \rangle  \\
 |\mathbf{8},S=0,I=0,I_z=0 \rangle \\
  |\mathbf{1},S=0,I=0,I_z=0 \rangle \\
 |\mathbf{27},S=0,I=1,I_z=0 \rangle\\
 |\mathbf{8},S=0,I=1,I_z=0 \rangle \end{array} \right) \nonumber\\
 &=&
 \mathbf{U}_{(\pi\pi)^0} \mathbf{A}_{(\pi\pi)^0}^{\rm bare}= \left(
\begin{array}{cccccc}
\sqrt{\frac{1}{3}} & -\sqrt{\frac{2}{3}}& 0& 0& 0& 0 \\
  \frac{1}{2\sqrt{15}}& \frac{1}{2\sqrt{30}}&
\frac{3\sqrt{3}}{2\sqrt{10}} & -\frac{\sqrt{3}}{2\sqrt{5}} &
-\frac{\sqrt{3}}{2\sqrt{5}} & 0\\
 -\sqrt{\frac{2}{5}} & -\sqrt{\frac{1}{5}}& \sqrt{\frac{1}{5}}
  & \frac{1}{\sqrt{10}} & \frac{1}{\sqrt{10}} & 0   \\
 \frac{1}{2}& \frac{1}{2\sqrt{2}}& \frac{1}{2\sqrt{2}}
  & \frac{1}{2} &\frac{1}{2} & 0\\
0& 0& 0& \sqrt{\frac{1}{5}} & -\sqrt{\frac{1}{5}}&
-\sqrt{\frac{3}{5}} \\
0& 0& 0& \sqrt{\frac{3}{10}} & -\sqrt{\frac{3}{10}}
  & \sqrt{\frac{2}{5}}
\end{array} \right)\mathbf{A}_{(\pi\pi)^0}^{\rm bare}\,,\label{app:su3pipi0}
\end{eqnarray}
where $\mathbf{A}_{(\overline{K\pi})^0}^{\rm bare},
\mathbf{A}_{({K\pi})^0}^{\rm bare}, \mathbf{A}_{({K\pi})^+}^{\rm
bare}, \mathbf{A}_{(\pi\pi)^+}^{\rm bare}$ and
$\mathbf{A}_{(\pi\pi)^0}^{\rm bare}$ have been defined in
Eqs.~(\ref{eq:subset1.2}), (\ref{eq:subset2.2}),
(\ref{eq:subset3.2}), (\ref{eq:subset4.2}) and
(\ref{eq:subset5.2}), respectively.


\begin{thebibliography}{99}

\bibitem{Buras:1985xv}
  A.~J.~Buras, J.~M.~Gerard and R.~Ruckl,
  Nucl.\ Phys.\ B {\bf 268}, 16 (1986).

\bibitem{Yeh:1997rq}
  T.~W.~Yeh and H.~n.~Li,
  Phys.\ Rev.\ D {\bf 56}, 1615 (1997)
  [arXiv:hep-ph/9701233].

\bibitem{KLS00}
Y.Y. Keum, H.-n.\ Li and A.I. Sanda,Phys.\ Rev.\ D {\bf 63},
054008, (2001) [arXiv:hep-ph/0004173].

\bibitem{Beneke:2001ev}
  M.~Beneke, G.~Buchalla, M.~Neubert and C.~T.~Sachrajda,
  Nucl.\ Phys.\ B {\bf 606}, 245 (2001)
  [arXiv:hep-ph/0104110].

\bibitem{Beneke:2003zv}
  M.~Beneke and M.~Neubert,
  Nucl.\ Phys.\ B {\bf 675}, 333 (2003)
  [arXiv:hep-ph/0308039].

\bibitem{Qiu:1988dn}
  J.~W.~Qiu,
  Phys.\ Rev.\ D {\bf 42}, 30 (1990).

\bibitem{Yang:1997er}
  K.~C.~Yang and H.~L.~Yu,
  Phys.\ Lett.\ B {\bf 430}, 186 (1998)
  [arXiv:hep-ph/9712366].

\bibitem{Abe:2001zi}
K.~Abe {\it et al.}  [BELLE Collaboration],
Phys.\ Rev.\ Lett.\  {\bf 88}, 052002 (2002)
[arXiv:hep-ex/0109021];
{\it ibid} {\bf 87}, 111801 (2001) [hep-ex/0104051].

\bibitem{Coan:2001ei}
T.E.~Coan {\it et al.}  [CLEO Collaboration],
Phys.\ Rev.\ Lett.\  {\bf 88}, 062001 (2002)
[arXiv:hep-ex/0110055].



\bibitem{Krokovny:2002pe}
  P.~Krokovny {\it et al.}  [Belle Collaboration],
  Phys.\ Rev.\ Lett.\  {\bf 89}, 231804 (2002)
  [arXiv:hep-ex/0207077];
{\it ibid}  {\bf 90}, 141802 (2003) [arXiv:hep-ex/0212066].

\bibitem{Aubert:2002vg}
  B.~Aubert {\it et al.}  [BABAR Collaboration],
  Phys.\ Rev.\ Lett.\  {\bf 90}, 181803 (2003)
  [arXiv:hep-ex/0211053];
  Phys.\ Rev.\ D {\bf 69}, 032004 (2004)
  [arXiv:hep-ex/0310028].

\bibitem{Schumann:2005ej}
  J.~Schumann {\it et al.}  [BELLE Collaboration],
  Phys.\ Rev.\ D {\bf 72}, 011103 (2005)
  [arXiv:hep-ex/0501013].

\bibitem{Cheng:1998kd}
  H.~Y.~Cheng and K.~C.~Yang,
  Phys.\ Rev.\ D {\bf 59}, 092004 (1999)
  [arXiv:hep-ph/9811249].

\bibitem{Yang}
  C.~K.~Chua, W.~S.~Hou and K.~C.~Yang,
  Phys.\ Rev.\ D {\bf 65}, 096007 (2002)
  [arXiv:hep-ph/0112148].

\bibitem{PDG} Particle Data Group,  S. Eidelman {\it et al.,} \pl B
{\bf 592}, 1 (2004).

\bibitem{Chua:2002wk}
C.~K.~Chua, W.~S.~Hou and K.~C.~Yang,
Mod.\ Phys.\ Lett.\ A {\bf 18}, 1763 (2003)
[arXiv:hep-ph/0210002].

\bibitem{Feldmann:1998vh}
T.~Feldmann, P.~Kroll and B.~Stech,
Phys.~Rev.~D {\bf 58}, 114006 (1998) [hep-ph/9802409];
Phys.\ Lett.\ {\bf B449}, 339 (1999) [hep-ph/9812269].

\bibitem{BraF}
V.M. Braun and I.E. Filyanov, Z.\ Phys.\ C {\bf 44} (1989) 157;
Z.\ Phys.\ C {\bf 48} (1990) 239.

\bibitem{Geshkenbein:qn}
B.~V.~Geshkenbein and M.~V.~Terentev,
Yad.\ Fiz.\  {\bf 40} (1984) 758 [Sov.\ J.\ Nucl.\ Phys.\  {\bf
40} (1984) 487].

\bibitem{Watson}  K. M. Watson, \textit{Phys. Rev.} \textbf{88}, 1163 (1952).

\bibitem{Suzuki:1999uc}
M.~Suzuki and L.~Wolfenstein,
Phys.\ Rev.\ D {\bf 60}, 074019 (1999) [hep-ph/9903477].

\bibitem{Smith:1998nu}
  C.~Smith,
  Eur.\ Phys.\ J.\ C {\bf 10}, 639 (1999)
  [arXiv:hep-ph/9808376].


\bibitem{Semenov:2003ne}
  S.~V.~Semenov,
  Phys.\ Atom.\ Nucl.\  {\bf 66}, 526 (2003)
  [Yad.\ Fiz.\  {\bf 66}, 553 (2003)].

\bibitem{Chen:1999nx}
  Y.~H.~Chen, H.~Y.~Cheng, B.~Tseng and K.~C.~Yang,
  Phys.\ Rev.\ D {\bf 60}, 094014 (1999)
  [arXiv:hep-ph/9903453].

\bibitem{Buchalla:1995vs}
  G.~Buchalla, A.~J.~Buras and M.~E.~Lautenbacher,
  Rev.\ Mod.\ Phys.\  {\bf 68}, 1125 (1996)
  [arXiv:hep-ph/9512380].

\bibitem{Ball:2004hn}
P.~Ball and R.~Zwicky,
Phys.\ Rev.\ D {\bf 71}, 014015 (2005) [arXiv:hep-ph/0406232];
  arXiv:hep-ph/0406261.

\bibitem{Colangelo:2003vg}
  P.~Colangelo and F.~De Fazio,
  Phys.\ Lett.\ B {\bf 570}, 180 (2003)
  [arXiv:hep-ph/0305140].

\bibitem{Kagan:2004uw}
A.~L.~Kagan,
Phys.\ Lett.\ B {\bf 601}, 151 (2004) [arXiv:hep-ph/0405134].


\bibitem{Yang:2005tv}
K.~C.~Yang,
Phys.\ Rev.\ D{\bf 72} (2005) 034009 [arXiv:hep-ph/0506040].

\bibitem{Cheng:2002wu}
  H.~Y.~Cheng,
  Eur.\ Phys.\ J.\ C {\bf 26}, 551 (2003)
  [arXiv:hep-ph/0202254].

\bibitem{Dai:1999cs}
  Y.~S.~Dai, D.~S.~Du, X.~Q.~Li, Z.~T.~Wei and B.~S.~Zou,
  Phys.\ Rev.\ D {\bf 60}, 014014 (1999)
  [arXiv:hep-ph/9903204].

\bibitem{Eeg:2001un}
  J.~O.~Eeg, S.~Fajfer and J.~Zupan,
  Phys.\ Rev.\ D {\bf 64}, 034010 (2001)
  [arXiv:hep-ph/0101215].

\bibitem{Ball:2003sc}
  P.~Ball and M.~Boglione,
  Phys.\ Rev.\ D {\bf 68}, 094006 (2003)
  [arXiv:hep-ph/0307337].




\bibitem{Braun:2004vf}
  V.~M.~Braun and A.~Lenz,
  Phys.\ Rev.\ D {\bf 70}, 074020 (2004)
  [arXiv:hep-ph/0407282].

\bibitem{Khodjamirian:2004ga}
A.~Khodjamirian, T.~Mannel and M.~Melcher,
Phys.\ Rev.\ D {\bf 70}, 094002 (2004) [arXiv:hep-ph/0407226].

\bibitem{Chernyak:1981zz}
  V.~L.~Chernyak and A.~R.~Zhitnitsky,
  Nucl.\ Phys.\ B {\bf 201}, 492 (1982)
  [Erratum-ibid.\ B {\bf 214}, 547 (1983)].

\bibitem{Chau:1982da}
  L.~L.~Chau,
  Phys.\ Rept.\  {\bf 95}, 1 (1983).

\bibitem{Chau:1986jb}
  L.~L.~Chau and H.~Y.~Cheng,
  Phys.\ Rev.\ Lett.\  {\bf 56}, 1655 (1986).

\bibitem{Chau:1987tk}
  L.~L.~Chau and H.~Y.~Cheng,
  Phys.\ Rev.\ D {\bf 36}, 137 (1987).

\bibitem{Rosner:1999xd}
  J.~L.~Rosner,
  Phys.\ Rev.\ D {\bf 60}, 114026 (1999)
  [arXiv:hep-ph/9905366].

\bibitem{Neubert:1997wb}
  M.~Neubert,
  Phys.\ Lett.\ B {\bf 424}, 152 (1998)
  [arXiv:hep-ph/9712224].

\bibitem{Chiang:2002mr}
  C.~W.~Chiang, Z.~Luo and J.~L.~Rosner,
  Phys.\ Rev.\ D {\bf 67}, 014001 (2003)
  [arXiv:hep-ph/0209272].

\bibitem{Gronau:1999zt}
  M.~Gronau,
  Phys.\ Rev.\ Lett.\  {\bf 83}, 4005 (1999)
  [arXiv:hep-ph/9908237].

\bibitem{Elaaoud:1999pj}
  E.~h.~El aaoud and A.~N.~Kamal,
  Int.\ J.\ Mod.\ Phys.\ A {\bf 15}, 4163 (2000)
  [arXiv:hep-ph/9910327].

\bibitem{Ablikim:2002ep}
  M.~Ablikim, D.~S.~Du and M.~Z.~Yang,
  Phys.\ Lett.\ B {\bf 536}, 34 (2002)
  [arXiv:hep-ph/0201168].

\bibitem{Lee}  T. D. Lee, \textit{Particle Physics and Introduction to Field
Theory}, Harwood Academic Publishers, 1988.

\bibitem{georgi}
H.~Georgi, \textit{Lie Algebras in Particle Physics,}
Addison-Wesley, 1982.


\end{thebibliography}
\end{document}